\documentclass[fleqn,usenatbib]{mnras}

\usepackage{newtxtext,newtxmath}

\usepackage[T1]{fontenc}

\DeclareRobustCommand{\VAN}[3]{#2}
\let\VANthebibliography\thebibliography
\def\thebibliography{\DeclareRobustCommand{\VAN}[3]{##3}\VANthebibliography}

\usepackage{graphicx}	
\usepackage{amsmath}	
\usepackage{tabularx}
\usepackage{adjustbox}  

\title[Polarization of hot Jupiters]{Polarization of hot Jupiter systems: a likely detection of stellar activity and a possible detection of planetary polarization}

\author[J. Bailey et al.]{Jeremy Bailey,$^{1}$\thanks{E-mail: j.bailey@unsw.edu.au (JB)}
Kimberly Bott,$^{2}$
Daniel V. Cotton,$^{3,4,5}$
Lucyna Kedziora-Chudczer,$^{5}$
\newauthor Jinglin Zhao,$^{1,6}$
Dag Evensberget,$^{5}$
Jonathan P. Marshall,$^{5,7}$
Duncan Wright,$^{5}$
P.W. Lucas$^{8}$
\\
$^{1}$School of Physics, University of New South Wales, Sydney, NSW 2052, Australia\\
$^{2}$Department of Earth and Planetary Science, University of California, Riverside, CA, 92521, USA\\
$^{3}$Anglo Australian Telescope, Australian National University, 418 Observatory Road, Coonabarabran, NSW 2357, Australia.\\
$^{4}$Western Sydney University, Locked Bag 1797, Penrith-South DC, NSW 1797, Australia.\\
$^{5}$Centre for Astrophysics, University of Southern Queensland, Toowoomba, Queensland. 4350. Australia.\\
$^{6}$Department of Astronomy \& Astrophysics, The Pennsylvania State University, 525 Davey Laboratory, University Park, PA 16802, USA.\\ 
$^{7}$Academia Sinica, Institute of Astronomy and Astrophysics, 11F Astronomy-Mathematics Building, NTU/AS campus,\\ No. 1, Section 4, Roosevelt Rd., Taipei 10617, Taiwan\\
$^{8}$Centre for Astrophysics, University of Hertfordshire, College Lane, Hatfield, AL10 9AB, UK.\\
}

\date{Accepted XXX. Received YYY; in original form ZZZ}

\pubyear{2020}

\begin{document}
\label{firstpage}
\pagerange{\pageref{firstpage}--\pageref{lastpage}}
\maketitle

\begin{abstract}
We present high-precision linear polarization observations of four bright hot Jupiter systems ($\tau$ Boo, HD 179949, HD 189733 and 51 Peg) and use the data to search for polarized reflected light from the planets. The data for 51 Peg are consistent with a reflected light polarization signal at about the level expected with 2.8$\sigma$ significance and a false alarm probability of 1.9 per cent. More data will be needed to confirm a detection of reflected light in this system. HD 189733 shows highly variable polarization that appears to be most likely the result of magnetic activity of the host star. This masks any polarization due to reflected light, but a polarization signal at the expected level of $\sim$20 ppm cannot be ruled out. $\tau$ Boo and HD 179949 show no evidence for polarization due to reflected light. The results are consistent with the idea that many hot Jupiters have low geometric albedos. Conclusive detection of polarized reflected light from hot Jupiters is likely to require further improvements in instrument sensitivity.
\end{abstract}

\begin{keywords}
polarization -- techniques: polarimetric -- planets and satellites: atmospheres -- stars: activity
\end{keywords}



\section{Introduction}

Observations of reflected light from hot Jupiters have the potential to provide a method of characterization of their atmospheres that complements other methods such as transmission spectroscopy \citep[e.g.][]{sing16}, and thermal emission \citep[e.g.][]{crouzet14}. Hot Jupiters have large radii and orbit close to their stars maximizing the amount of flux incident on their disk, and hence enhancing the chances of those photons being reflected. The reflected light signals in the combined light of the star and the planet are predicted to be small \citep{seager00}, and most detections so far have been by space-based photometry \citep[e.g.][]{demory11}. However, reflected light is also expected to be polarized, and polarization is a differential measurement that can be made to very high precision from ground-based telescopes. Stellar polarimeters capable of measuring to parts-per-million (ppm) levels have been built \citep{hough06,wiktorowicz08,piirola14,wiktorowicz15a,bailey15,bailey20}.

Ground-based polarization measurements should therefore be capable of detecting the expected polarization variation around the orbital cycle. The polarization in the combined light of the planet and star is predicted to be at levels from a few ppm to tens of ppm \citep{seager00,bailey18} in favourable cases. Previous reports of variable polarization from HD~189733b \citep{berdyugina08,berdyugina11} with amplitudes of 100-200 ppm (much larger than that predicted) were not confirmed by subsequent studies \citep{wiktorowicz09,wiktorowicz15,bott16}, while only upper limits on reflected light polarization have been reported for $\tau$ Boo and 55 Cnc \citep{lucas09} and WASP-18 \citep{bott18}. 

Here we report new polarization observations of four of the brightest hot Jupiter systems made with the High Precision Polarimetric Instrument \citep[HIPPI,][]{bailey15} and its successor \citep[HIPPI-2,][]{bailey20} on the 3.9-m Anglo-Australian Telescope.

This paper is structured as follows. Section \ref{sec:objects} describes the four hot Jupiter systems and discusses the reflected light polarization signals that might be expected. Section \ref{sec:obs} presents the polarization observations. Section \ref{sec:discuss} discusses the results and considers a number of possible sources of polarization. The conclusions are presented in section \ref{sec:conclude}.

\begin{table*}
\caption{Planet and Star Properties.}
\setlength{\extrarowheight}{5pt}
\begin{tabular}{ |l|ccccl| }
\hline
\hline
 & $\tau$ Bootis b & HD 179949b & HD 189733b & 51 Pegasi b & References \\
\hline
\hline
\textbf{Planet} &&&&&\\
Mass (M$_{\rm Jup}$) & $5.90^{+0.35}_{-0.20}$ & 0.98$\pm0.04$ & $1.144^{+0.057}_{-0.056}$ & $0.476^{+0.032}_{-0.031}$ & 1,2,3,4\\
Radius (R$_{\rm Jup}$) & 1.21 & 1.22 & $1.138^{+0.027}_{-0.027}$ & $1.9^{+0.3}_{-0.3}$ & 5,5,3,6 \\
$a$ (AU) & 0.0462 & 0.0443 & $0.03120^{+0.00027}_{-0.00037}$ & 0.0527 & 7,8,9,8 \\
Inclination (deg) &  $45^{+3}_{-4}$ & 67.7$\pm4.3$ & $85.51^{+0.10}_{-0.05}$ & 70.0 - 82.2 & 1,2,9,4 \\ 
Orbital Period (d) & 3.312463 & 3.092514 & 2.21857312 & 4.2307869 & 10,2,9,4\\
Epoch (MJD) & 46956.416 & 51001.016 & 53988.30339 & 56326.4314 & 10,2,9,4 \\
Geometric Albedo & <0.12 &  &  $\begin{matrix}  0.4\pm0.12$ $(\rm blue)\\[-5pt] <0.12$ $(\rm green)\\ \end{matrix}$  & 0.5 & 11,12,6 \\
\hline
\textbf{Star} &&&&&\\
$m_V$ & 4.49 & 6.237 & 7.67 & 5.46 & 13,13,14,13 \\
Spectral Type & F7~IV-V & F8~V & K2~V & G2~IV & 13,13,14,13 \\
Activity Index -- $\log{(R^\prime_{\rm HK})}$ & $-$4.67 & $-$4.77 & $-$4.50 & $-$5.11  & 15 \\
Mean Magnetic Field (G) & 0.9 -- 3.9  & 2.6 -- 3.7 & 32 -- 42 & & 16,17,18 \\
Distance (pc) & 15.6 & 27.5 & 19.8 & 15.5 & 19,20,20,20 \\
\hline
\hline
\end{tabular}
\\
\vspace{0.5cm}
{\bf References}   1. \citet{lockwood14}, 2. \citet{brogi14}, 3. \citet{torres08}, 4. \citet{birkby17}, 5. \citet{sestovic18}, 6. \citet{martins15}, 7. \citet{butler97}, 8. \citet{butler06}, 9. \citet{triaud09}, 10. \citet{donati08}, 11. \citet{hoeijmakers18}, 12. \citet{evans13}, 13. \citet{valenti05}, 14. \citet{gray03}, 15. \citet{cantomartins11}, 16. \citet{mengel16}, 17. \citet{fares12}, 18. \citet{fares17}, 18. \citet{vanleeuwen07}, 20. \citet{gaia18}
\label{tab:properties}
\end{table*}

\section{Hot Jupiter Systems}
\label{sec:objects}

The hot Jupiter systems studied in this project have been chosen because their host stars are bright and their orbital periods are short. The planet and star properties are listed in Table \ref{tab:properties}. One other system (WASP-18) has also been observed as part of this project and the results are described elsewhere \citep{bott18}. The four exoplanets are all in orbits of less than a few days and have radii akin to Jupiter's.  Only HD~189733b transits its star, providing a reliable determination of its radius. Tau~Bo\"otis~b has a much greater mass (higher density) than the other planets presented here, and 51~Pegasi~b has a notably lower mass (lower density). 

The stellar environments of these systems are also important for polarized light observations, as stellar activity can produce a polarized light signal \citep{cotton17b, bott18, cotton19a}. Table \ref{tab:properties} includes information on the stellar activity and magnetic fields. All of the host stars of the planets examined in this paper are FGK stars: they are to some degree, Sun-like. Two of these host stars, $\tau$~Boo and 51~Peg, may not be dwarfs like our Sun, but may be sub-giants (type IV).  Relatively little information is available on intrinsic polarization in sub-giants, but the best current data \citep{bailey10, cotton16a, cotton17b} suggest a similarity with dwarfs (type~V). 

\subsection{\texorpdfstring{$\tau$ Bo\"otis}{tau Bootis}}
\label{sec:tauboo}

High-resolution cross-correlation observations of the thermal emission from the planet have been used to determine the inclination and detect molecular species (CO and H$_2$O) in this non-transiting planet \citep{brogi12,rodler12,lockwood14}. Having one of the brightest host stars for a hot Jupiter system, there have been many attempts to search for reflected light from the planet leading to upper limits on the geometric albedo in the range 0.3 to 0.4 \citep{charbonneau99, lucas09, rodler10, rodler12}. More recently \citet{hoeijmakers18} have reported a 3$\sigma$ upper limit on the geometric albedo of 0.12 from 400-700 nm.

$\tau$~Boo is a well known magnetically active star that has been extensively studied using Zeeman Doppler Imaging \citep{mengel16,jeffers18}. The star's rotation is synchronized with the orbital period of the planet. The magnetic field shows cyclic polarity reversals over a 240 day period analogous to the solar activity cycle \citep{jeffers18}. The magnetic activity confuses attempts to detect reflected light by photometric methods. Periodic variations seen by the MOST satellite \citep{walker08} were interpreted as due to an active spot on the stellar surface.

\subsection{HD 179949} 
\label{sec:hd179949}
HD~179949b does not transit its star. As in $\tau$~Boo~b, high spectral resolution cross correlation techniques have been used to detect molecular species (CO and H$_2$O) and constrain the inclination and mass \citep{brogi14}. The infrared phase curve shows indications of relatively poor heat redistribution compared to other hot Jupiters \citep{cowan07}.  

HD~179949 sometimes shows chromospheric activity synchronized with the planetary orbit, but at other epochs a stellar rotational period of 7 days is seen \citep{shkolnik03, shkolnik05, shkolnik08}. Spectropolarimetry shows a weak magnetic field of a few Gauss \citep{fares12} with differential rotation (7.62 days at equator, 10.3 days at the pole).

\subsection{HD 189733}
\label{sec:hd189733}
As a transiting planet orbiting a bright star, HD~189733b is one of the best studied of all known hot Jupiter systems.  Data has been obtained from radio, X-ray, infrared, and visible wavelengths in transit, radial velocity cross-correlation, secondary eclipse and polarized and non-polarized reflection, to paint a relatively complete portrait of the world and its environment.

Several molecules have been detected in the atmosphere of HD~189733b including CO and H$_2$O \citep{dekok13,birkby13,crouzet14,mccullough14,alonsofloriano19}. Sodium D-line absorption is also detected in the transmission spectrum \citep{redfield08}.

Although HD~189733b is expected to be tidally locked, it seems to have good heat transport to the nightside of the planet.  The emissions temperature from the nightside is ``only" a few hundred degrees cooler than the dayside \citep{wakeford15}.  The world is dynamic with high winds \citep{seidel20, louden15} and an offset hotspot shifted 21.8$\pm$1.5 degrees east of the substellar point \citep{knutson07, majeau12}.

Transit observations show a depth that increases at blue wavelengths \citep{pont13} indicating Rayleigh-like scattering from small dust grains. However, \citet{mccullough14} has suggested that a clear atmosphere could also fit the data, with the effect of star spots on the host star causing the wavelength dependence.   \citet{evans13} report a detection of reflected light at blue wavelengths with a geometric albedo  of $0.4_{-0.11}^{+0.12}$ short of 450~nm, and $0.0_{-0.10}^{+0.12}$ longward.

HD~189733 is an active K dwarf classified as a BY Draconis variable (showing variations due to star spots). A range of photometric rotation periods including 11.8 days \citep{hebrard06}, 13.4 days \citep{winn07}, and 11.953 days \citep{henry08} have been reported. Differential rotation ranging from 11.94 days at the equator to 16.53 days at the pole has been inferred from spectropolarimetry \citep{fares10}.
The mean magnetic field ranges from 32 to 42 G and is mostly toroidal dominated \citep{fares17}. The magnetic field is the strongest of the four stars included in this project.  Flares from the star have been shown to drive variations in atmospheric escape from the planet \citep{desEtangs12}.

\subsection{51 Pegasi} 
51~Pegasi~b was the first exoplanet orbiting a Sun-like star to be discovered \citep{mayor95}. It does not transit, but high-resolution cross-correlation methods have detected water absorption \citep{birkby17} and constrained the orbital inclination and mass of the planet. A detection of reflected light at 3$\sigma$ significance has been reported using HARPS spectroscopy \citep{martins15}, with the signal requiring an albedo of 0.5, and a planet radius of 1.9 $R_{\rm Jup}$. 

51~Peg is the least active of the four host stars included in this project as shown by its very low chromospheric activity index (see Table \ref{tab:properties}) and low soft X-ray flux \citep{poppenhager09}.

\bigbreak

\subsection{Expected Polarization Levels}

The polarization levels expected for reflected light from hot Jupiters have been investigated by \citet{bailey18} using the Versatile Software for Transfer of Atmospheric Radiation \citep[\textsc{vstar},][]{bailey12} code modified to include polarization using the Vector Linearized Discrete Ordinate Radiative Transfer (\textsc{vlidort}) code of \citet{spurr06}. Using an atmospheric model for HD~189733b that fits the observed thermal emission (secondary eclipse) and transmission (primary transit) spectra, a reflected light polarization amplitude of 27 ppm was calculated for the combined light of the star and planet. This was obtained for the most favourable case of optically thick Rayleigh scattering clouds. A more realistic cloud model resulted in a polarization amplitude of 20 ppm, and a geometric albedo close to that observed by \citet{evans13}. The results are consistent with calculations by \citet{lucas09}, \citet{buenzli09} and \citet{kopparla16}\footnote{after correction of a scaling error as described in \citet{kopparla18}}.

The optically thick Rayleigh scattering cloud models lead to geometric albedos for the planet of $\sim$0.7 and maximum polarizations for the planet of $\sim$35 percent limited by the occurrence of multiple scattering. The fractional polarization of the planetary light can be increased by reducing the cloud optical depth or making the cloud particles more absorbing, but this also reduces the geometric albedo and reduces the polarization signal seen in the combined light of the planet and star. We note that geometric albedos for hot Jupiters seen in space photometry of secondary eclipses or phase curves are much lower than 0.7. While reflected light albedos as high as $\sim$0.3 are seen in some cases such as Kepler-7b \citep[e.g.][]{demory11} most hot Jupiters show lower geometric albedos \citep{bailey14,angerhausen15,wong20}. These albedo measurements are mostly made with Kepler and TESS in relatively red bands, and it is possible that the planets become more reflective at bluer wavelengths, as suggested by the \citet{evans13} reflected light measurement for HD 189733b.

The three non-transiting planets included in this study are not sufficiently well characterized to allow a detailed model as used for HD~189733b. However, the most important factors that determine polarization levels are the value $(R_P/a)^2$ that determines the fraction of the star's light intercepted by the planet, and the cloud properties. Thus in Table \ref{tab:predict} we estimate the maximum polarization amplitude expected for the four planets, by using the model of \citet{bailey18} and scaling the result according to $(R_P/a)^2$. The figure given here for HD~189733b is slightly different to that given in \citet{bailey18} due to different adopted values for $R_P$ and $a$ in Table \ref{tab:properties}. We don't have direct measurements of the planet radius ($R_p$) for non-transiting planets. The radii listed in Table \ref{tab:properties} for $\tau$~Boo~b and HD~179949b are estimated using the empirical relationships given by \citet{sestovic18}, and take account of the effect of stellar irradiation on the inflation of the planet. The value of 1.9 $R_{\rm Jup}$ for 51 Peg is that from \citet{martins15} used to fit their reflected light detection. This is much higher than that given by the \citet{sestovic18} empirical relationship, which is 1.26 $R_{\rm Jup}$. Values for both these radii are given in Table \ref{tab:predict}.

\begin{table}
    \centering
    \caption{Predicted maximum polarization amplitudes for Hot Jupiter systems}
    \begin{tabular}{|c|c|c|c|}
    \hline
        Planet & $R_P$ & $(R_P/a)^2$ & Pol (ppm)  \\  \hline
        $\tau$ Boo b & 1.21 &  0.000156  &  13 \\
        HD 179949b & 1.22 & 0.000173  &  14 \\
        HD 189733b & 1.138 & 0.000304  &  25 \\
        51 Peg b & 1.9 & 0.000297 &  24 \\
        51 Peg b & 1.26 & 0.000118 &  11 \\
        \hline
    \end{tabular}
    \label{tab:predict}
\end{table}

The values listed in Table \ref{tab:predict} are for a wavelength of 440 nm and represent the best case possible, with optically thick clouds. The polarization will scale as $R_P^2$ for other radii. If the planet has no clouds, reflected light and polarization can still be produced by Rayleigh scattering from molecules. However, since light can then be more easily absorbed before it is scattered the geometric albedo and observed polarization are reduced. A clear atmosphere model of HD 189733b gave a polarization amplitude of 7 ppm \citep{bailey18}. However a similar model for the hotter planet, WASP-18b \citep{bott18} gave a polarization amplitude of only 0.17 ppm, because of the increased importance of other opacity sources such as H$^-$ and TiO absorption.

\begin{table*}
\caption{TP and PA Calibrations by Run.}
\centering
\tabcolsep 3.9 pt
\begin{tabular}{|lccccr||cccccc|rr||ccccccccc|r|}
\hline
\hline
Run & f/ &  Inst  & Fil & Mod  & \multicolumn{1}{c||}{Ap} &   \multicolumn{6}{c|}{LP Std. Obs.} & \multicolumn{1}{c}{$Q/I$} & \multicolumn{1}{c||}{$U/I$} &  \multicolumn{9}{c|}{HP Std. Obs.} & S.D.\\
&&&&Era&\multicolumn{1}{c||}{(\arcsec)}& A & B & C & D & E & F &\multicolumn{1}{c}{(ppm)} & \multicolumn{1}{c||}{(ppm)} & A & B & C & D & E & F & G & H & I & \multicolumn{1}{c|}{($^{\circ}$)} \\
\hline
\hline
2014AUG & 8 & H & 500SP & E1 & 6.6                              & 3 & 0 & 0 & 0 & 0 & 1 & $-$42.1 $\pm$ 2.1 & $-$37.3 $\pm$ 2.1 & 1 & 0 & 0 & 0 & 2 & 0 & 0 & 4 & 0 & 0.28 \\
2015MAY & 8 & H & Clear & E1 & 6.6                              & 0 & 0 & 1 & 0 & 0 & 3 & $-$37.6 $\pm$ 2.0 & $-$2.3 $\pm$ 2.1  & 0 & 0 & 0 & 0 & 4 & 1 & 0 & 0 & 0 & 0.17 \\
2015MAY$^{C1}$ & 8 & H & 500SP & E1 & 6.6                       & 1 & 0 & 1 & 0 & 0 & 2 & $-$39.0 $\pm$ 2.4 & 4.8 $\pm$ 2.3     & 0 & 0 & 0 & 0 & 4 & 1 & 0 & 0 & 0 & 0.17 \\
2015JUN$^{C1}$ & 8 & H & 500SP & E1 & 6.6                       & 1 & 0 & 1 & 0 & 0 & 2 & $-$39.0 $\pm$ 2.4 & 4.8 $\pm$ 2.3     & 0 & 0 & 0 & 0 & 1 & 1 & 0 & 0 & 0 & 0.14 \\
2015OCT & 8 & H & Clear & E1 & 6.6                              & 5 & 0 & 6 & 0 & 0 & 0 & $-$45.4 $\pm$ 0.8 & $-$0.5 $\pm$ 0.7  & 1 & 0 & 0 & 0 & 0 & 2 & 0 & 2 & 0 & 0.24 \\
2016FEB & 8 & H & Clear & E2 & 6.6                              & 0 & 0 & 4 & 0 & 1 & 0 & $-$14.5 $\pm$ 0.8 & 2.2 $\pm$ 0.8     & 0 & 1 & 0 & 0 & 1 & 0 & 0 & 0 & 0 & 0.29 \\
2017JUN$^{C2}$ & 8 & H & Clear & E2 & 6.6                       & 6 & 0 & 5 & 3 & 2 & 0 & $-$14.9 $\pm$ 1.1 & 1.0 $\pm$ 1.1     & 0 & 0 & 0 & 0 & 2 & 1 & 1 & 0 & 0 & 1.11 \\
2017JUN$^{C3}$ & 8 & H & 500SP & E2 & 6.6                       & 2 & 0 & 3 & 2 & 2 & 0 & $-$10.0 $\pm$ 1.7 & $-$0.4 $\pm$ 1.7  & 0 & 0 & 0 & 0 & 2 & 1 & 1 & 0 & 0 & 1.11 \\
2017AUG$^{C2}$ & 8 & H & Clear & E2 & 6.6                       & 6 & 0 & 5 & 3 & 2 & 0 & $-$14.9 $\pm$ 1.1 & 1.0 $\pm$ 1.1     & 0 & 0 & 0 & 0 & 1 & 1 & 0 & 1 & 0 & 0.53 \\
2017AUG$^{C3}$ & 8 & H & 500SP & E2 & 6.6                       & 2 & 0 & 3 & 2 & 2 & 0 & $-$10.0 $\pm$ 1.7 & $-$0.4 $\pm$ 1.7  & 0 & 0 & 0 & 0 & 1 & 1 & 0 & 1 & 0 & 0.53 \\
2018FEB-B & 15 & H2 & Clear & E3 & 9.2                          & 0 & 0 & 2 & 1 & 0 & 0 & $-$181.1 $\pm$ 1.1 & 20.8 $\pm$ 1.1   & 0 & 1 & 0 & 0 & 0 & 0 & 0 & 0 & 0 & \multicolumn{1}{c|}{--}\\
2018MAR & 8$^{\times2}$ & H2 & Clear & E3 & 8.6                 & 0 & 0 & 2 & 0 & 3 & 0 & 123.1 $\pm$ 1.2 & $-$12.2 $\pm$ 1.3   & 1 & 0 & 0 & 1 & 1 & 0 & 0 & 1 & 0 & 0.26 \\
2018JUL$^{C4}$ & 8$^{\times2}$ & H2 & Clear & E4 & 11.9         & 3 & 3 & 2 & 2 & 2 & 2 & $-$10.1 $\pm$ 0.9 & 3.8 $\pm$ 0.9     & 0 & 0 & 0 & 0 & 1 & 1 & 1 & 1 & 1 & 1.55 \\
2018JUL & 8$^{\times2}$ & H2 & 500SP & E4 & 11.9                & 2 & 2 & 2 & 2 & 1 & 2 & 0.6 $\pm$ 1.9 & 18.4 $\pm$ 1.4        & 0 & 0 & 0 & 0 & 1 & 1 & 1 & 1 & 1$^R$ & 1.55 \\
2018AUG$^{C4}$ & 8$^{\times2}$ & H2 & Clear & E5--E7* & 11.9    & 3 & 3 & 2 & 2 & 2 & 2 & $-$10.1 $\pm$ 0.9 & 3.8 $\pm$ 0.9     & 0 & 0 & 0 & 0 & 3 & 0 & 3 & 5 & 0 & 0.86 \\
\hline
\hline
\end{tabular}
\begin{flushleft}
Notes: \\
Runs are named for the month in which they were begun. A new run begins every time the instrument is re-installed on the telescope. \\
LP standards: A:~HD~2151, B:~HD~10700, C:~HD~48915, D:~HD~102647, E:~HD~102870, F:~HD~140573. \\
HP standards: A:~HD~23512, B:~HD~80558, C:~HD~84810, D:~HD~111613, E:~HD~147084, F:~HD~154445, G:~HD~160529, H:~HD~187929, I:~HD~203532. \\
$^C$ TP standard observations in the same filter were combined for runs: 2015MAY \& 2015JUN (C1), 2017JUN \& 2017AUG (C2 and C3), 2018JUL \& 2018AUG (C4). \\
$^{\times2}$~indicates use of the $\times2$ negative achromatic (Barlow) lens to adjust the focus for HIPPI-2, which is designed for f/16. \\
$^R$ 2018JUL 500SP measurements were rotated 5.6$^\circ$ from the g$^\prime$/Clear calibration based on the difference between HD~203532 observations in g$^\prime$ and 500SP. \\
* The modulator evolution during this run encompassed three eras; E5: JD 2458346.5 -- 2458354.5, E6: 2458354.5 -- 2458358.5, E7: 2458358.5 -- 2458364.5.
\end{flushleft}
\label{tab:runs}
\end{table*}

\begin{table*}
    \centering
    \caption{Linear Polarization Observations of $\tau$ Boo}
    \tabcolsep 10.5 pt
    \begin{tabular}{|ccc|rr|ccccc|}
    \hline
    \hline
    HMJD & Phase & (Range) & \multicolumn{1}{c}{$Q/I$} & \multicolumn{1}{c|}{$U/I$} & Run & n & Exp & $\lambda_{\rm eff}$ & Eff. \\
     &  &  & \multicolumn{1}{c}{(ppm)} & \multicolumn{1}{c|}{(ppm)} & & & (s) & (nm) & (\%) \\
    \hline
    \hline
57165.43196 & 0.00151 & (.99055-.01365) &    6.3 $\pm$   3.9 &   18.4 $\pm$   3.9 & 2015MAY & 3 & 3840 & 485.7 & 85.2 \\
57168.58721 & 0.95405 & (.94739-.96220) &    8.1 $\pm$   4.7 &   12.5 $\pm$   4.7 & 2015MAY & 2 & 2560 & 487.4 & 85.5 \\
57446.64223 & 0.89612 & (.88599-.90690) &    6.8 $\pm$   3.7 &   14.7 $\pm$   3.7 & 2016FEB & 3 & 3840 & 486.5 & 83.7 \\
57447.69000 & 0.21244 & (.19653-.22851) &    8.0 $\pm$   3.3 &    1.5 $\pm$   3.3 & 2016FEB & 4 & 5120 & 485.1 & 83.3 \\
57448.68425 & 0.51259 & (.50203-.52661) &   14.9 $\pm$   3.6 &   11.9 $\pm$   3.7 & 2016FEB & 3 & 3840 & 485.0 & 83.3 \\
\hline
57926.43447 & 0.74068 & (.73158-.75228) &   17.7 $\pm$   4.3 &   $-$6.9 $\pm$   4.3 & 2017JUN & 2 & 3840 & 484.7 & 83.2 \\
57928.41595 & 0.33887 & (.32147-.35102) &    9.3 $\pm$   4.0 &   $-$2.4 $\pm$   4.2 & 2017JUN & 3 & 3840 & 484.7 & 83.2 \\
57929.39578 & 0.63467 & (.62424-.64618) &   13.6 $\pm$   4.7 &   $-$9.9 $\pm$   4.6 & 2017JUN & 2 & 4480 & 484.6 & 83.2 \\
57930.39641 & 0.93675 & (.92478-.95085) &   10.9 $\pm$   4.3 &   $-$4.1 $\pm$   4.3 & 2017JUN & 2 & 3840 & 484.7 & 83.2 \\
57933.44273 & 0.85641 & (.84728-.86769) &   26.0 $\pm$   5.0 &   $-$8.7 $\pm$   5.1 & 2017JUN & 2 & 3840 & 485.2 & 83.3 \\
57934.41956 & 0.15130 & (.14282-.16173) &   19.4 $\pm$   4.4 &  $-$15.8 $\pm$   4.5 & 2017JUN & 2 & 3840 & 484.8 & 83.2 \\
57935.42021 & 0.45339 & (.44498-.46358) &   14.6 $\pm$   4.3 &   $-$3.0 $\pm$   4.3 & 2017JUN & 2 & 3840 & 484.9 & 83.3 \\
57936.43504 & 0.75976 & (.75127-.77025) &   15.4 $\pm$   4.3 &   $-$9.1 $\pm$   4.3 & 2017JUN & 2 & 3840 & 485.3 & 83.4 \\
57939.44604 & 0.66875 & (.65977-.67899) &   20.0 $\pm$   4.5 &    4.8 $\pm$   4.5 & 2017JUN & 2 & 3840 & 485.9 & 83.6 \\
\hline
57975.37115 & 0.51418 & (.50808-.52130) &   $-$0.1 $\pm$   4.9 &    3.5 $\pm$   5.0 & 2017AUG & 2 & 2080 & 487.2 & 83.8 \\
57977.37172 & 0.11813 & (.11262-.12505) &    7.3 $\pm$   4.9 &  $-$12.7 $\pm$   5.1 & 2017AUG & 2 & 2080 & 487.7 & 84.0 \\
57978.36883 & 0.41915 & (.41368-.42750) &    7.1 $\pm$   5.0 &    3.3 $\pm$   5.1 & 2017AUG & 2 & 2400 & 487.7 & 84.0 \\
57980.36877 & 0.02291 & (.01704-.02991) &   12.8 $\pm$   5.3 &  $-$18.8 $\pm$   5.3 & 2017AUG & 2 & 2080 & 488.1 & 84.0 \\
57981.37365 & 0.32628 & (.32037-.33267) &    9.5 $\pm$   5.5 &  $-$11.9 $\pm$   5.4 & 2017AUG & 2 & 2000 & 488.9 & 84.2 \\
57984.36863 & 0.23043 & (.22564-.23623) &   10.0 $\pm$   6.7 &   $-$2.1 $\pm$   6.4 & 2017AUG & 2 & 1600 & 489.2 & 84.3 \\
\hline
58152.72803 & 0.05648 & (.05019-.06351) &    2.1 $\pm$   3.9 &   19.6 $\pm$   3.9 & 2018FEB-B & 3 & 2400 & 485.2 & 78.7 \\
58200.66345 & 0.52771 & (.52035-.53573) &   13.5 $\pm$   3.8 &    6.0 $\pm$   3.7 & 2018MAR & 2 & 2560 & 487.1 & 79.9 \\
58201.65342 & 0.82657 & (.81614-.83808) &    6.9 $\pm$   2.9 &    2.5 $\pm$   2.9 & 2018MAR & 3 & 3840 & 487.2 & 79.9 \\
58203.65504 & 0.43084 & (.42031-.44296) &    7.0 $\pm$   2.9 &   $-$8.7 $\pm$   2.9 & 2018MAR & 3 & 3840 & 487.2 & 79.9 \\
58204.63863 & 0.72778 & (.72097-.73586) &   13.9 $\pm$   3.7 &   11.6 $\pm$   3.7 & 2018MAR & 2 & 2560 & 487.2 & 79.9 \\
58204.67637 & 0.73917 & (.73674-.74282) &   $-$5.5 $\pm$   5.2 &   $-$2.1 $\pm$   5.0 & 2018MAR & 1 & 1280 & 487.2 & 79.9 \\
58205.65168 & 0.03361 & (.02359-.04561) &   11.9 $\pm$   2.9 &    6.6 $\pm$   2.9 & 2018MAR & 3 & 3840 & 487.2 & 79.9 \\
58206.62240 & 0.32666 & (.31730-.33742) &    9.6 $\pm$   3.0 &    2.0 $\pm$   2.9 & 2018MAR & 3 & 3840 & 487.3 & 79.9 \\
58207.63729 & 0.63305 & (.62350-.64389) &   11.8 $\pm$   3.0 &    9.8 $\pm$   2.9 & 2018MAR & 3 & 3840 & 487.2 & 79.9 \\
58210.64703 & 0.54166 & (.53563-.54876) &   10.8 $\pm$   3.8 &    1.6 $\pm$   3.7 & 2018MAR & 2 & 2560 & 487.2 & 79.9 \\
58211.66718 & 0.84963 & (.83973-.86071) &    3.0 $\pm$   3.0 &    6.7 $\pm$   3.0 & 2018MAR & 3 & 3840 & 487.4 & 80.0 \\
58212.63779 & 0.14265 & (.13360-.15355) &   15.5 $\pm$   3.0 &    5.4 $\pm$   2.9 & 2018MAR & 3 & 3840 & 487.2 & 79.9 \\
58213.65073 & 0.44845 & (.43904-.45890) &   10.1 $\pm$   2.9 &    4.9 $\pm$   3.0 & 2018MAR & 3 & 3840 & 487.3 & 79.9 \\
58214.65266 & 0.75092 & (.74166-.76165) &   18.1 $\pm$   3.0 &    7.3 $\pm$   3.0 & 2018MAR & 3 & 3840 & 487.3 & 79.9 \\
58215.66582 & 0.05678 & (.04751-.06743) &   13.4 $\pm$   3.1 &   12.0 $\pm$   3.0 & 2018MAR & 3 & 3840 & 487.6 & 80.0 \\
58216.64566 & 0.35258 & (.34296-.36315) &   11.0 $\pm$   3.0 &    8.4 $\pm$   2.9 & 2018MAR & 3 & 3840 & 487.3 & 79.9 \\
\hline
58309.43526 & 0.36485 & (.35814-.37316) &  $-$15.3 $\pm$   3.5 &   12.9 $\pm$   3.5 & 2018JUL & 2 & 2560 & 488.4 & 79.7 \\
58310.42566 & 0.66384 & (.65450-.67453) &    5.4 $\pm$   2.9 &    8.9 $\pm$   2.9 & 2018JUL & 3 & 3840 & 488.3 & 79.7 \\
58311.41112 & 0.96134 & (.95251-.97171) &    1.0 $\pm$   3.0 &   $-$4.7 $\pm$   2.9 & 2018JUL & 3 & 3840 & 487.9 & 79.5 \\
58312.42113 & 0.26626 & (.25684-.27686) &   $-$2.7 $\pm$   3.0 &   10.9 $\pm$   3.0 & 2018JUL & 3 & 3840 & 488.3 & 79.7 \\
58313.39982 & 0.56171 & (.55308-.57177) &    3.9 $\pm$   2.9 &    2.6 $\pm$   2.8 & 2018JUL & 3 & 3840 & 487.8 & 79.5 \\
58314.37806 & 0.85703 & (.85145-.86414) &   $-$4.9 $\pm$   3.7 &   $-$1.3 $\pm$   3.7 & 2018JUL & 2 & 2560 & 487.3 & 79.3 \\
58315.40071 & 0.16576 & (.16007-.17262) &    0.9 $\pm$   3.5 &   $-$3.9 $\pm$   3.5 & 2018JUL & 2 & 2560 & 487.9 & 79.6 \\
58316.40748 & 0.46970 & (.46244-.47848) &   $-$8.3 $\pm$   3.2 &   15.8 $\pm$   3.1 & 2018JUL & 3 & 3200 & 488.2 & 79.7 \\
58317.41642 & 0.77428 & (.76573-.78439) &    0.2 $\pm$   2.9 &   18.2 $\pm$   2.9 & 2018JUL & 3 & 3840 & 488.6 & 79.8 \\
58318.40503 & 0.07274 & (.06709-.07979) &   $-$7.0 $\pm$   3.6 &    8.2 $\pm$   3.6 & 2018JUL & 2 & 2560 & 488.2 & 79.7 \\
58319.37141 & 0.36448 & (.36149-.37001) &   $-$3.6 $\pm$   5.0 &   $-$3.5 $\pm$   5.2 & 2018JUL & 1 & 1600 & 487.4 & 79.4 \\
58321.39084 & 0.97412 & (.96545-.98411) &    5.2 $\pm$   2.9 &   12.6 $\pm$   2.9 & 2018JUL & 3 & 3840 & 488.1 & 79.6 \\
\hline
58347.36874 & 0.81660 & (.81421-.81996) &   $-$0.2 $\pm$   5.6 &    6.9 $\pm$   5.4 & 2018AUG & 1 & 1280 & 490.8 & 74.2 \\
\hline
\hline
    \end{tabular}
    \label{tab:tauboo}
\end{table*}

\section{Observations}
\label{sec:obs}

We have made high precision polarimetric observations of four hot Jupiter exoplanet systems with the 3.9-m Anglo-Australian Telescope (AAT), located at Siding Spring Observatory in Australia. Between August 2014 and August 2017 these observations were made with the HIgh Precision Polarimetric Instrument \citep[HIPPI,][]{bailey15}. From February 2018 to August 2018 the observations were made with HIPPI-2 \citep{bailey20}. Altogether the observations span eleven observing runs. The two instruments use Hamamatsu H10720-210 photomultuplier tube modules as detectors. These have ultrabialkali photocathodes \citep{nakamura10} providing a quantum efficiency of 43~per~cent at 400~nm. The instruments use a Boulder Nonlinear Systems (BNS) Ferro-electric Liquid Crystal (FLC) modulator operating at 500 Hz. The precision of the instruments is determined by summing in quadrature the internal measurement precision, which depends on exposure time, and an error introduced by uncertainties in the centering of the star in the aperture. This ``positioning error'', which represents the ultimate precision, is determined empirically from many observations of bright stars as described by \citet{bailey20}.

\begin{table*}
    \centering
    \caption{Linear Polarization Observations of HD~179949}
    \tabcolsep 10.5 pt
    \begin{tabular}{|ccc|rr|ccccc|}
    \hline
    \hline
    HMJD & Phase & (Range) & \multicolumn{1}{c}{$Q/I$} & \multicolumn{1}{c|}{$U/I$} & Run & n & Exp & $\lambda_{\rm eff}$ & Eff. \\
     &  &  & \multicolumn{1}{c}{(ppm)} & \multicolumn{1}{c|}{(ppm)} & & & (s) & (nm) & (\%) \\
    \hline
    \hline
57164.61564 & 0.07089 & (.05818-.08447) &  $-$24.3 $\pm$   \phantom{0}6.3 &   $-$8.7 $\pm$   \phantom{0}6.3 & 2015MAY & 3 & 3840 & 485.2 & 84.9 \\
57165.64521 & 0.40382 & (.39355-.41422) &  $-$36.5 $\pm$   \phantom{0}7.3 &  $-$12.3 $\pm$   \phantom{0}7.4 & 2015MAY & 2 & 2480 & 484.5 & 84.8 \\
57166.64914 & 0.72845 & (.71930-.74069) &  $-$10.5 $\pm$   \phantom{0}8.8 &    2.3 $\pm$   \phantom{0}9.2 & 2015MAY & 2 & 2560 & 484.3 & 84.8 \\
57167.70102 & 0.06859 & (.05424-.09048) &  $-$29.0 $\pm$   \phantom{0}6.3 &   $-$5.5 $\pm$   \phantom{0}6.2 & 2015MAY & 3 & 3840 & 483.9 & 84.7 \\
57168.64003 & 0.37223 & (.36533-.38080) &  $-$23.0 $\pm$   \phantom{0}6.8 &  $-$12.9 $\pm$   \phantom{0}6.8 & 2015MAY & 2 & 2560 & 484.4 & 84.8 \\
\hline
57926.52579 & 0.44331 & (.43800-.45471) &   $-$9.4 $\pm$   \phantom{0}7.9 &  $-$19.7 $\pm$   \phantom{0}8.1 & 2017JUN & 1 & 3200 & 485.3 & 83.2 \\
57929.47552 & 0.39714 & (.39184-.40657) &  $-$21.6 $\pm$   \phantom{0}8.6 &  $-$28.0 $\pm$   \phantom{0}9.0 & 2017JUN & 1 & 2880 & 487.1 & 83.7 \\
57930.52203 & 0.73554 & (.72366-.74999) &  $-$22.2 $\pm$   \phantom{0}6.0 &  $-$23.8 $\pm$   \phantom{0}6.0 & 2017JUN & 2 & 5120 & 485.2 & 83.1 \\
57933.54548 & 0.71321 & (.70078-.72679) &  $-$16.3 $\pm$   \phantom{0}7.1 &  $-$22.3 $\pm$   \phantom{0}7.0 & 2017JUN & 2 & 4240 & 484.5 & 83.0 \\
57934.51617 & 0.02709 & (.01578-.04158) &    3.8 $\pm$   \phantom{0}6.6 &  $-$13.1 $\pm$   \phantom{0}6.3 & 2017JUN & 2 & 5120 & 485.0 & 83.2 \\
57935.51465 & 0.34996 & (.33861-.36385) &   $-$9.2 $\pm$   \phantom{0}5.8 &  $-$24.1 $\pm$   \phantom{0}5.8 & 2017JUN & 2 & 5120 & 485.0 & 83.1 \\
57936.53201 & 0.67894 & (.66734-.69298) &    4.7 $\pm$   \phantom{0}5.7 &  $-$17.1 $\pm$   \phantom{0}5.7 & 2017JUN & 2 & 5120 & 484.5 & 83.0 \\
\hline
57978.62008 & 0.28860 & (.27953-.29990) &    5.9 $\pm$   \phantom{0}6.5 &  $-$26.7 $\pm$   \phantom{0}6.4 & 2017AUG & 2 & 3840 & 485.5 & 83.3 \\
57979.63685 & 0.61738 & (.60360-.63418) &  $-$20.3 $\pm$   \phantom{0}5.5 &  $-$17.1 $\pm$   \phantom{0}5.5 & 2017AUG & 4 & 5120 & 486.5 & 83.5 \\
57980.58438 & 0.92378 & (.91328-.93699) &  $-$34.2 $\pm$   \phantom{0}7.4 &  $-$29.8 $\pm$   \phantom{0}7.6 & 2017AUG & 2 & 4480 & 484.7 & 83.1 \\
57981.58505 & 0.24736 & (.23594-.26149) &  $-$12.6 $\pm$   \phantom{0}6.0 &  $-$20.5 $\pm$   \phantom{0}5.9 & 2017AUG & 2 & 5120 & 484.7 & 83.1 \\
57982.66216 & 0.59565 & (.58679-.60740) &  $-$21.4 $\pm$   \phantom{0}7.7 &  $-$20.5 $\pm$   \phantom{0}7.9 & 2017AUG & 2 & 3840 & 488.4 & 84.0 \\
57983.62539 & 0.90712 & (.90081-.92298) &  $-$23.1 $\pm$  10.6 &  $-$18.4 $\pm$  11.7 & 2017AUG & 1 & 3520 & 486.2 & 83.4 \\
57984.58410 & 0.21713 & (.20542-.23077) &  $-$32.4 $\pm$   \phantom{0}7.1 &  $-$21.9 $\pm$   \phantom{0}7.1 & 2017AUG & 2 & 5120 & 485.0 & 83.2 \\
57985.40440 & 0.48239 & (.47338-.49355) &  $-$17.4 $\pm$   \phantom{0}6.8 &   $-$9.2 $\pm$   \phantom{0}6.8 & 2017AUG & 2 & 3840 & 484.4 & 83.0 \\
\hline
58200.72356 & 0.10831 & (.09695-.12057) &  $-$26.0 $\pm$   \phantom{0}5.6 &  $-$18.7 $\pm$   \phantom{0}5.6 & 2018MAR & 3 & 3840 & 490.0 & 80.6 \\
58201.73361 & 0.43493 & (.42086-.45037) &  $-$21.1 $\pm$   \phantom{0}5.0 &   $-$6.3 $\pm$   \phantom{0}5.1 & 2018MAR & 4 & 4800 & 489.4 & 80.5 \\
58203.72488 & 0.07883 & (.06814-.09298) &  $-$40.3 $\pm$   \phantom{0}6.5 &   $-$6.6 $\pm$   \phantom{0}6.3 & 2018MAR & 3 & 3840 & 489.4 & 80.5 \\
58203.77702 & 0.09569 & (.09412-.09789) &  $-$43.1 $\pm$  13.2 &  $-$29.7 $\pm$  13.2 & 2018MAR & 1 & 640 & 487.5 & 79.9 \\
58204.70806 & 0.39675 & (.39072-.40391) &  $-$35.0 $\pm$   \phantom{0}6.9 &  $-$24.9 $\pm$   \phantom{0}6.6 & 2018MAR & 2 & 2560 & 490.1 & 80.7 \\
58204.75542 & 0.41206 & (.40457-.42093) &  $-$13.2 $\pm$   \phantom{0}6.7 &   $-$5.6 $\pm$   \phantom{0}6.6 & 2018MAR & 2 & 2560 & 488.0 & 80.1 \\
58216.69977 & 0.27441 & (.26773-.28232) &  $-$35.3 $\pm$   \phantom{0}6.6 &  $-$10.7 $\pm$   \phantom{0}6.7 & 2018MAR & 2 & 2560 & 488.9 & 80.3 \\
\hline
58309.53090 & 0.29242 & (.28555-.30050) &  $-$28.6 $\pm$   \phantom{0}6.1 &   $-$8.1 $\pm$   \phantom{0}6.0 & 2018JUL & 2 & 2560 & 486.9 & 79.1 \\
58310.51808 & 0.61164 & (.60173-.62365) &  $-$31.6 $\pm$   \phantom{0}5.2 &   $-$4.0 $\pm$   \phantom{0}5.2 & 2018JUL & 3 & 3840 & 487.1 & 79.2 \\
58311.51122 & 0.93278 & (.91955-.94738) &  $-$23.2 $\pm$   \phantom{0}4.7 &  $-$14.1 $\pm$   \phantom{0}4.7 & 2018JUL & 4 & 5120 & 487.2 & 79.3 \\
58312.50916 & 0.25547 & (.24153-.27004) &  $-$33.6 $\pm$   \phantom{0}4.5 &  $-$14.7 $\pm$   \phantom{0}4.5 & 2018JUL & 4 & 5120 & 487.2 & 79.3 \\
58313.48209 & 0.57008 & (.55729-.58430) &  $-$42.4 $\pm$   \phantom{0}4.3 &  $-$13.4 $\pm$   \phantom{0}4.4 & 2018JUL & 4 & 5120 & 487.6 & 79.4 \\
58314.44024 & 0.87991 & (.87045-.89053) &  $-$31.6 $\pm$   \phantom{0}5.2 &  $-$10.4 $\pm$   \phantom{0}5.1 & 2018JUL & 3 & 3840 & 488.9 & 79.8 \\
58315.51780 & 0.22835 & (.21918-.23862) &  $-$37.3 $\pm$   \phantom{0}5.0 &   $-$3.9 $\pm$   \phantom{0}5.0 & 2018JUL & 3 & 3840 & 486.9 & 79.2 \\
58316.69247 & 0.60820 & (.59912-.61891) &  $-$43.1 $\pm$   \phantom{0}5.2 &   $-$5.5 $\pm$   \phantom{0}5.3 & 2018JUL & 3 & 3840 & 488.1 & 79.5 \\
58317.68588 & 0.92943 & (.92019-.94004) &  $-$29.7 $\pm$   \phantom{0}5.1 &  $-$10.9 $\pm$   \phantom{0}5.1 & 2018JUL & 3 & 3840 & 487.9 & 79.5 \\
\hline
\hline
    \end{tabular}
    \label{tab:hd179949}
\end{table*}

\begin{table*}
    \centering
    \caption{Linear Polarization Observations of HD~189733}
    \tabcolsep 10.5 pt
    \begin{tabular}{|ccc|rr|ccccc|}
    \hline
    \hline
    HMJD & Phase & (Range) & \multicolumn{1}{c}{$Q/I$} & \multicolumn{1}{c|}{$U/I$} & Run & n & Exp & $\lambda_{\rm eff}$ & Eff. \\
     &  &  & \multicolumn{1}{c}{(ppm)} & \multicolumn{1}{c|}{(ppm)} & & & (s) & (nm) & (\%) \\
    \hline
    \hline
56897.51655 & 0.29920 & (.27215-.33105) &   46.6 $\pm$  13.8 &   25.6 $\pm$  13.7 & 2014AUG & 3 & 7680 & 450.0 & 85.7 \\
56898.47180 & 0.72977 & (.70072-.76316) &   17.3 $\pm$  12.3 &   15.4 $\pm$  12.6 & 2014AUG & 3 & 7680 & 450.0 & 85.7 \\
56899.50048 & 0.19344 & (.16291-.22435) &   53.4 $\pm$  12.8 &   16.4 $\pm$  13.0 & 2014AUG & 3 & 7680 & 450.0 & 85.7 \\
56900.46649 & 0.62886 & (.60374-.65953) &   93.7 $\pm$  12.6 &   26.3 $\pm$  12.6 & 2014AUG & 3 & 7040 & 449.9 & 85.6 \\
57165.74238 & 0.19935 & (.17103-.22798) &   42.8 $\pm$  13.3 &   24.8 $\pm$  13.3 & 2015MAY & 3 & 7680 & 449.9 & 85.7 \\
57168.72494 & 0.54371 & (.51819-.57446) &   70.6 $\pm$  12.4 &   33.8 $\pm$  12.4 & 2015MAY & 3 & 7680 & 449.9 & 85.7 \\
57199.70770 & 0.50889 & (.49309-.52856) &   11.4 $\pm$  15.1 &   65.8 $\pm$  15.0 & 2015JUN & 2 & 5120 & 450.0 & 85.7 \\
\hline
57929.65623 & 0.52600 & (.49624-.54992) &    0.9 $\pm$  17.0 &    7.9 $\pm$  15.7 & 2017JUN & 2 & 5920 & 449.8 & 82.2 \\
57930.58274 & 0.94362 & (.93654-.96303) &  $-$13.9 $\pm$  19.5 &    9.7 $\pm$  19.6 & 2017JUN & 1 & 3840 & 450.5 & 82.6 \\
57933.64304 & 0.32302 & (.29826-.35319) &   $-$6.4 $\pm$  15.4 &   $-$2.0 $\pm$  15.8 & 2017JUN & 3 & 7680 & 449.9 & 82.3 \\
57934.65618 & 0.77968 & (.75335-.80816) &  $-$24.3 $\pm$  14.4 &   27.8 $\pm$  14.1 & 2017JUN & 3 & 7680 & 449.9 & 82.3 \\
57935.66986 & 0.23659 & (.21247-.26556) &  $-$22.3 $\pm$  12.1 &    9.4 $\pm$  12.3 & 2017JUN & 3 & 7680 & 449.9 & 82.3 \\
57936.65479 & 0.68054 & (.65686-.71864) &   81.2 $\pm$  16.0 &  $-$18.3 $\pm$  16.0 & 2017JUN & 2 & 5440 & 449.9 & 82.3 \\
\hline
57978.46128 & 0.52440 & (.51353-.55326) &    3.0 $\pm$  23.1 &   34.7 $\pm$  21.7 & 2017AUG & 1 & 3360 & 450.3 & 82.5 \\
57979.49546 & 0.99055 & (.96005-.01624) &   26.2 $\pm$  16.2 &    0.5 $\pm$  16.4 & 2017AUG & 4 & 5440 & 450.1 & 82.4 \\
57980.48009 & 0.43436 & (.40962-.46540) &   10.9 $\pm$  15.6 &    7.7 $\pm$  15.9 & 2017AUG & 3 & 7680 & 450.1 & 82.4 \\
57981.48441 & 0.88705 & (.86095-.91521) &    0.6 $\pm$  12.7 &   21.4 $\pm$  12.8 & 2017AUG & 3 & 7680 & 450.0 & 82.4 \\
57982.56503 & 0.37412 & (.34972-.40422) &   37.3 $\pm$  14.8 &   44.9 $\pm$  15.2 & 2017AUG & 3 & 7680 & 450.2 & 82.4 \\
57983.53833 & 0.81283 & (.79399-.84209) &   73.8 $\pm$  19.9 &   52.1 $\pm$  21.1 & 2017AUG & 2 & 5440 & 449.9 & 82.3 \\
57984.48259 & 0.23845 & (.21271-.26684) &   49.3 $\pm$  16.7 &   34.2 $\pm$  16.8 & 2017AUG & 3 & 7680 & 450.0 & 82.3 \\
57985.48803 & 0.69164 & (.67041-.72100) &   30.5 $\pm$  14.8 &   60.4 $\pm$  15.0 & 2017AUG & 3 & 7040 & 449.9 & 82.3 \\
\hline
58309.61470 & 0.78854 & (.76303-.81462) &    4.9 $\pm$  12.0 &    8.5 $\pm$  11.9 & 2018JUL & 5 & 6400 & 451.7 & 76.3 \\
58310.62730 & 0.24497 & (.22227-.27022) &   20.8 $\pm$  12.1 &   36.2 $\pm$  12.0 & 2018JUL & 5 & 6400 & 451.7 & 76.2 \\
58311.60970 & 0.68777 & (.66437-.71337) &   $-$1.0 $\pm$  12.8 &   36.1 $\pm$  12.6 & 2018JUL & 5 & 6400 & 451.7 & 76.2 \\
58312.60954 & 0.13844 & (.11504-.16349) &  $-$30.1 $\pm$  12.6 &   $-$4.7 $\pm$  12.8 & 2018JUL & 5 & 6400 & 451.7 & 76.2 \\
58313.60193 & 0.58575 & (.55845-.61499) &    2.9 $\pm$  10.8 &   36.2 $\pm$  10.5 & 2018JUL & 6 & 7680 & 451.8 & 76.3 \\
58314.60251 & 0.03675 & (.01065-.06521) &   $-$2.5 $\pm$  10.9 &   27.9 $\pm$  10.7 & 2018JUL & 6 & 7680 & 451.8 & 76.3 \\
58315.61101 & 0.49132 & (.46492-.51934) &    1.3 $\pm$  10.9 &   73.4 $\pm$  10.9 & 2018JUL & 6 & 7680 & 451.7 & 76.3 \\
58316.59832 & 0.93634 & (.91090-.96415) &    7.7 $\pm$  11.0 &   49.2 $\pm$  10.9 & 2018JUL & 6 & 7680 & 451.7 & 76.3 \\
58317.59980 & 0.38775 & (.36529-.41180) &   29.2 $\pm$  11.9 &   29.9 $\pm$  11.9 & 2018JUL & 5 & 6400 & 451.7 & 76.2 \\
58318.60248 & 0.83970 & (.83605-.84980) &   22.3 $\pm$  22.5 &  $-$12.7 $\pm$  23.1 & 2018JUL & 1 & 1910 & 451.6 & 76.2 \\
\hline
\hline
    \end{tabular}
    \label{tab:hd189733}
\end{table*}

\begin{table*}
    \centering
    \caption{Linear Polarization Observations of 51~Peg}
    \tabcolsep 10.5 pt
    \begin{tabular}{|ccc|rr|ccccc|}
    \hline
    \hline
    HMJD & Phase & (Range) & \multicolumn{1}{c}{$Q/I$} & \multicolumn{1}{c|}{$U/I$} & Run & n & Exp & $\lambda_{\rm eff}$ & Eff. \\
     &  &  & \multicolumn{1}{c}{(ppm)} & \multicolumn{1}{c|}{(ppm)} & & & (s) & (nm) & (\%) \\
    \hline
    \hline
57309.48699 & 0.35762 & (.34817-.36903) &  $-$10.7 $\pm$   5.0 &   $-$5.1 $\pm$   5.1 & 2015OCT & 2 & 5120 & 491.9 & 86.0 \\
57310.46707 & 0.58928 & (.57989-.60021) &   $-$7.7 $\pm$   4.9 &   $-$8.3 $\pm$   4.9 & 2015OCT & 2 & 5120 & 491.9 & 86.0 \\
57311.46637 & 0.82548 & (.81677-.83587) &   $-$7.7 $\pm$   4.8 &  $-$10.5 $\pm$   4.8 & 2015OCT & 2 & 5120 & 491.9 & 86.0 \\
\hline
57929.74735 & 0.96401 & (.95534-.97447) &   $-$3.1 $\pm$   4.9 &  $-$19.1 $\pm$   5.0 & 2017JUN & 2 & 5120 & 492.2 & 84.7 \\
57933.74803 & 0.90962 & (.90070-.91989) &    7.3 $\pm$   6.2 &  $-$27.5 $\pm$   6.1 & 2017JUN & 2 & 5120 & 492.0 & 84.6 \\
57934.75340 & 0.14725 & (.13911-.15764) &    5.3 $\pm$   5.0 &  $-$21.0 $\pm$   5.0 & 2017JUN & 2 & 5120 & 492.0 & 84.6 \\
57935.76985 & 0.38750 & (.37896-.39791) &   $-$4.2 $\pm$   4.8 &  $-$25.9 $\pm$   4.8 & 2017JUN & 2 & 5120 & 491.9 & 84.6 \\
57938.76946 & 0.09649 & (.08807-.10661) &   $-$5.7 $\pm$   4.9 &   $-$8.4 $\pm$   4.9 & 2017JUN & 2 & 5120 & 491.9 & 84.6 \\
57939.77223 & 0.33351 & (.31533-.34599) &   12.9 $\pm$   5.0 &  $-$31.5 $\pm$   4.9 & 2017JUN & 3 & 4640 & 492.2 & 84.7 \\
\hline
57978.69238 & 0.53278 & (.52448-.54314) &    3.8 $\pm$   4.9 &  $-$19.1 $\pm$   4.9 & 2017AUG & 2 & 5120 & 492.3 & 84.7 \\
57980.66510 & 0.99906 & (.99079-.01030) &   10.7 $\pm$   6.9 &  $-$39.8 $\pm$   6.9 & 2017AUG & 2 & 5120 & 492.0 & 84.6 \\
57981.66984 & 0.23654 & (.22732-.24786) &  $-$11.3 $\pm$   5.2 &  $-$31.8 $\pm$   5.3 & 2017AUG & 2 & 5120 & 492.1 & 84.6 \\
57982.72823 & 0.48671 & (.48040-.49445) &   $-$4.7 $\pm$   6.4 &  $-$12.0 $\pm$   6.4 & 2017AUG & 2 & 3540 & 494.1 & 85.1 \\
57984.66504 & 0.94450 & (.93598-.95471) &   $-$9.1 $\pm$   6.1 &   $-$6.4 $\pm$   6.2 & 2017AUG & 2 & 5120 & 492.1 & 84.7 \\
\hline
58309.71760 & 0.77478 & (.76437-.78580) &  $-$27.9 $\pm$   3.5 &   $-$8.0 $\pm$   3.4 & 2018JUL & 4 & 5120 & 494.5 & 81.3 \\
58310.72413 & 0.01269 & (.00334-.02301) &  $-$17.3 $\pm$   3.5 &  $-$18.7 $\pm$   3.4 & 2018JUL & 4 & 5120 & 494.4 & 81.3 \\
58311.70903 & 0.24548 & (.23580-.25602) &  $-$22.0 $\pm$   3.8 &  $-$15.5 $\pm$   3.8 & 2018JUL & 4 & 5120 & 494.5 & 81.3 \\
58312.74544 & 0.49045 & (.48101-.50123) &  $-$13.7 $\pm$   3.6 &   $-$1.5 $\pm$   3.6 & 2018JUL & 4 & 5120 & 494.3 & 81.3 \\
58313.72456 & 0.72188 & (.71238-.73232) &  $-$19.1 $\pm$   3.4 &  $-$12.1 $\pm$   3.3 & 2018JUL & 4 & 5120 & 494.3 & 81.3 \\
58314.70609 & 0.95387 & (.94449-.96427) &  $-$15.8 $\pm$   3.5 &   $-$5.3 $\pm$   3.5 & 2018JUL & 4 & 5120 & 494.4 & 81.3 \\
58315.74187 & 0.19869 & (.18943-.20906) &  $-$13.2 $\pm$   3.5 &  $-$10.4 $\pm$   3.4 & 2018JUL & 4 & 5120 & 494.3 & 81.3 \\
58316.76302 & 0.44005 & (.43100-.45038) &  $-$24.8 $\pm$   3.5 &   $-$1.3 $\pm$   3.5 & 2018JUL & 4 & 5120 & 494.6 & 81.4 \\
58317.75718 & 0.67504 & (.66584-.68549) &  $-$18.0 $\pm$   3.5 &  $-$19.5 $\pm$   3.5 & 2018JUL & 4 & 5120 & 494.6 & 81.4 \\
58319.72699 & 0.14063 & (.13118-.15102) &  $-$11.3 $\pm$   3.5 &  $-$10.3 $\pm$   3.5 & 2018JUL & 4 & 5120 & 494.3 & 81.3 \\
58320.77338 & 0.38795 & (.38375-.39387) &  $-$19.4 $\pm$   5.3 &  $-$18.4 $\pm$   5.0 & 2018JUL & 2 & 2560 & 495.0 & 81.5 \\
58321.75585 & 0.62017 & (.61065-.63103) &   $-$4.1 $\pm$   3.4 &   $-$9.8 $\pm$   3.4 & 2018JUL & 4 & 5120 & 494.8 & 81.4 \\
\hline
58346.58004 & 0.48768 & (.47877-.49715) &   $-$3.6 $\pm$   3.9 &   $-$6.5 $\pm$   3.9 & 2018AUG & 4 & 5120 & 495.4 & 75.9 \\
58347.60855 & 0.73078 & (.72431-.73805) &  $-$18.5 $\pm$   4.3 &  $-$28.8 $\pm$   4.3 & 2018AUG & 3 & 3840 & 494.5 & 75.6 \\
58348.65013 & 0.97698 & (.97059-.98433) &   $-$3.1 $\pm$   4.4 &   $-$2.6 $\pm$   4.4 & 2018AUG & 3 & 3840 & 494.3 & 75.4 \\
58362.51975 & 0.25524 & (.25033-.26146) &   $-$0.8 $\pm$   6.4 &   $-$6.2 $\pm$   5.8 & 2018AUG & 2 & 3080 & 496.0 & 67.8 \\
58363.53352 & 0.49485 & (.48611-.50464) &    0.7 $\pm$   4.4 &  $-$24.4 $\pm$   4.5 & 2018AUG & 4 & 5120 & 495.4 & 67.6 \\ 

\hline
\hline
    \end{tabular}
    \label{tab:51peg}
\end{table*}

In addition to some of the most sensitive searches for exoplanets \citep{bott16, bott18} and studies of the polarisation in active dwarfs \citep{cotton17b, cotton19a}, HIPPI has been successfully used for a range of science programs including surveys of polarisation in bright stars \citep{cotton16a}, the first detection of polarisation due to rapid rotation in hot stars \citep{cotton17a}; reflection from the photospheres of a binary star \citep{bailey19}; studies of debris disc systems \citep{cotton17b, marshall20}, the interstellar medium \citep{cotton17b, cotton19b} and hot dust \citep{marshall16}. HIPPI-2 has recently been used in the study of reflected light in binary systems \citep{bailey19, cotton20c}, the rapidly rotating star $\alpha$~Oph \citep{bailey20b}, the red supergiant Betelgeuse \citep{cotton20b} and the polluted white dwarf G29-38 \citep{cotton20a}.

Observations of three of the systems -- $\tau$~Boo, 51~Peg and HD~179949 -- were made with no filter (Clear) giving a wavelength range from about 350 -- 750 nm with the response peak near 400 nm. For HD~189733, which is the reddest of the four objects with a K2 V host star, we used  a 500~nm short-pass filter (denoted 500SP) removing the redder part of this range. The effective wavelengths, taking account of the colour of the stars, are about 450~nm for the HD 189733 observations and 485-495 nm for the other targets. Full details of the instrument's wavelength response can be found in \citet{bailey20}. The positioning errors of HIPPI and HIPPI-2 in Clear are 4.7~ppm and 2.9~ppm respectively, in 500SP the figures are 7.5~ppm and 6.2~ppm. Other features of the two instruments, and their configuration, resulted in minor performance differences between runs. 

The bandpasses for the two instruments are slightly different; the collimating lens used with HIPPI attenuates some blue wavelengths, as does the Barlow lens that was used with HIPPI-2 at the AAT f/8 Cassegrain focus (but not the f/15 Cassegrain focus used for the 2018FEB-B observing run). Airmass, which we calculate to two decimal places, also has a small effect on the bandpass.  Effective wavelengths are calculated for each individual observation taking account of these effects and of the colour of the stars observed. As reported in detail by \citet{bailey20} the performance of the BNS modulator drifted over time -- the wavelength of peak modulation efficiency increased. This required its operation to be divided up into performance ``eras'' where the modulation efficiency curve was re-calibrated through observations of high polarization standards. It is important to note that in this case, although the effective wavelength ($\lambda_{\rm eff}$) of the instrument is not affected, the wavelengths most sensitive to polarization are. 

A small telescope polarization, TP, due to the telescope mirrors, shifts the zero-point offset of the observations. Because the polarization signals we are looking for are very small, the calibration of TP is critical for these observations. In the first instance TP is determined and corrected for by reference to the straight mean of several observations of low polarization standard stars, a summary of which are given in Table \ref{tab:runs}. Similarly, the position angle (PA) is calibrated by reference to literature measurements of high polarization standards -- listed in \citet{bailey20} -- which are also summarised in Table \ref{tab:runs}. TP calibrations are made in the same band as the observations, while the PA calibration is initially made with observations in SDSS g$^{\prime}$ and Clear, with corrections applied for other bands, where necessary, based on a smaller number of observations. A minor software glitch sometimes induced a 0.3$^{\circ}$ error in PA for HIPPI-2 observations -- this is largely inconsequential for observations of low polarization objects such as those we are interested in here.

While the standards are sufficiently bright to be observed with precision at any lunar phase, the fainter exoplanet target observations were largely restricted to dark (HD~189733) or at least grey (51~Peg, HD~179949) sky conditions. The standard observing procedure for HIPPI-class instruments involves taking a single sky (S) measurement adjacent to each target (T) measurement at each of the four position angles, $PA=0, 45, 90, 135^{\circ}$, in the pattern TSSTTSST \citep{bailey15, bailey17, bailey20}. Occasionally when it was deemed necessary, because of either brighter moonlight or non-ideal weather conditions, target measurements were bracketed between two skies. With HIPPI, which used the Cassegrain rotator of the AAT, the target was always re-centred for each PA. HIPPI-2's instrument rotator is more accurate, and allows for centering just once, which is always carried out at PA~=~0$^{\circ}$; this was the procedure where the aperture was larger than 10$\arcsec$. Smaller apertures were sometimes chosen to reduce the sky background contribution, in this case the object was also re-centred at each PA. 

The reduction procedure involves combining measurements at 0 and 90$^{\circ}$ and 45 and 135$^{\circ}$ to cancel instrumental polarization. For this reason individual observations are kept to not much more than an hour of dwell time at most, which minimises differences between the paired measurements due to airmass or sky condition changes. Multiple back-to-back observations were therefore required for the exposure to achieve the desired precision. In latter runs, each new observation in a set was begun at PA~=~0$^{\circ}$, while particularly for HIPPI observations the second in a set was sometimes begun at PA~=~135$^{\circ}$. In Tables \ref{tab:tauboo}, \ref{tab:hd179949}, \ref{tab:hd189733}, and \ref{tab:51peg} we report the the nightly error-weighted mean of the multiple observations (after calibration) for each of the four systems, together with orbital phase calculated from the ephemeris data in Table \ref{tab:properties}.  The results for HD 189733 given in Table \ref{tab:hd189733} include the observations previously published by \citet{bott16}. As the observations have been reprocessed using the methods described in \citet{bailey20} the numbers are slightly different from those given originally.

\section{Results and Discussion}
\label{sec:discuss}

\begin{figure*}
\includegraphics[width=8.8cm]{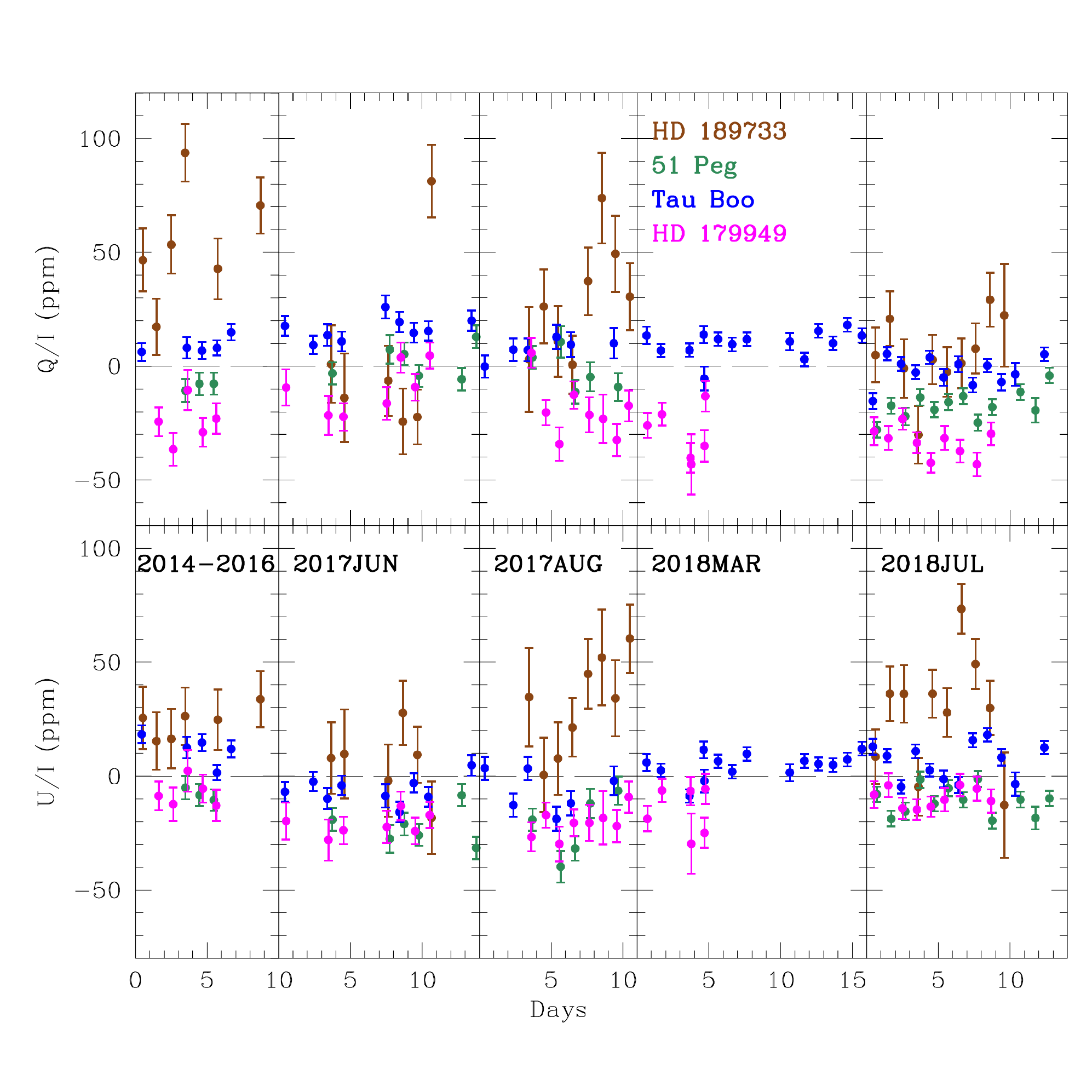}
\includegraphics[width=8.8cm]{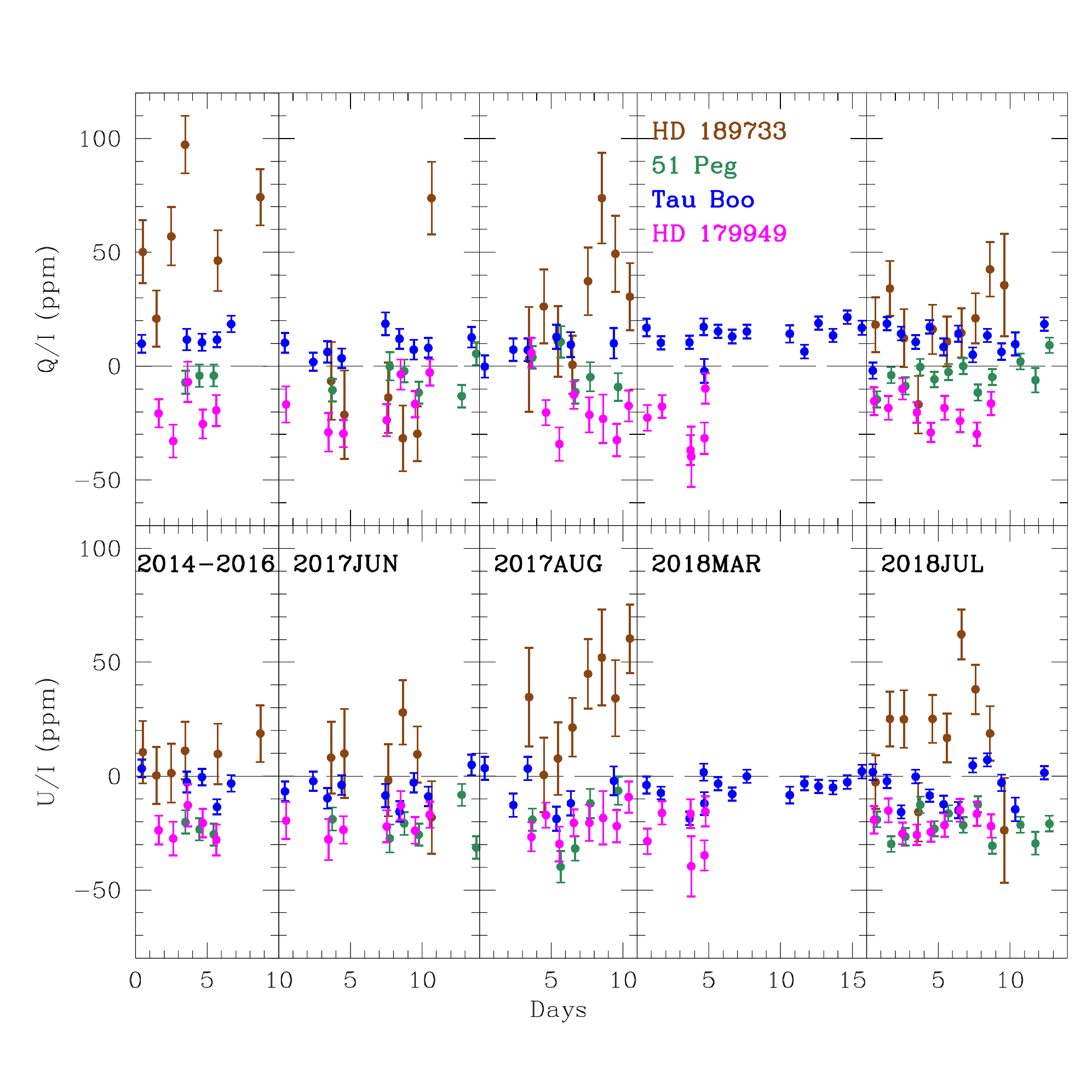}
\caption{Polarization data on hot Jupiters for the four longest contiguous observing runs (2017JUN, 2017AUG, 2018MAR and 2018JUL), as well as some data from short runs over the 2014 to 2016 period. The left panel shows data according to the standard reduction as listed in Tables \ref{tab:tauboo} to \ref{tab:51peg}. The right panel shows data with corrected zero point offsets as listed in Table \ref{tab:offsets}. }
\label{fig:runs}
\end{figure*}

The polarization data for the four longest contiguous observing runs and some of the shorter runs during 2014--2016 are shown in Fig. \ref{fig:runs}. The left panel shows the data according to the standard reduction procedures as listed in Tables \ref{tab:tauboo} to \ref{tab:51peg}. It can be seen that there are large variations between runs in the polarization of HD 189733, and smaller differences (at $\sim$ 10 ppm level) between runs for some of the other objects. Such small offsets can be due to a number of possible effects. The observations are made with two different instruments and observing procedures have changed over time. HIPPI observations were made by centering the object in the aperture at each of the four observation angles (0, 45, 90 and 135 degrees) whereas HIPPI-2 observations, which used larger apertures and a more accurate rotator, centered the object only at the 0 degree angle. These different procedures could result in different offsets due to the positioning instrumental effects described by \citet{bailey20}. The telescope polarization changes between runs and was particularly large for the 2018FEB-B and 2018MAR runs. The set of unpolarized standard stars used to fix the zero point were different for different runs. While the reduction assumes these stars to have zero polarization they may well have polarization at levels that are significant for high-precision studies like this. We cannot rule out the possibility that there is some variability in these low polarization standards \citep{lucas09}. 

The fact that the offsets in polarization between runs are almost the same for $\tau$~Boo, 51~Peg and HD~179949 leads us to conclude that these offsets are due to instrumental and calibration issues as described above. By taking the average value of Q/I and U/I across runs for these three objects, and taking the 2017AUG run as the reference, we derived the offset corrections listed in Table \ref{tab:offsets}. These offsets are applied to the polarization values plotted in the right panel of Fig. \ref{fig:runs} and for the data used in subsequent analysis.

The much larger changes seen in HD~189733 are too large to be accounted for in the same way and indicate real polarization variability of this object.

\begin{table}
    \centering
    \caption{Zero point offset corrections}
    \begin{tabular}{lrr}
         Run(s) & Offset (Q/I) & Offset (U/I) \\
         \hline
         2014--2016  &  +3.6 &  $-$15.1 \\
         2017JUN &  $-$7.4  &   +0.2   \\
         2017AUG  &  0.0  &  0.0  \\
         2018FEB-B, 2018MAR  &  +3.4  & $-$9.9  \\
         2018JUL    &  +13.3  &  $-$11.1 \\
         2018AUG    &   +3.0  &   $-$8.1 \\
         \hline
    \end{tabular}
    \label{tab:offsets}
\end{table}

\begin{figure*}
    \centering
    \includegraphics[width=8.8cm]{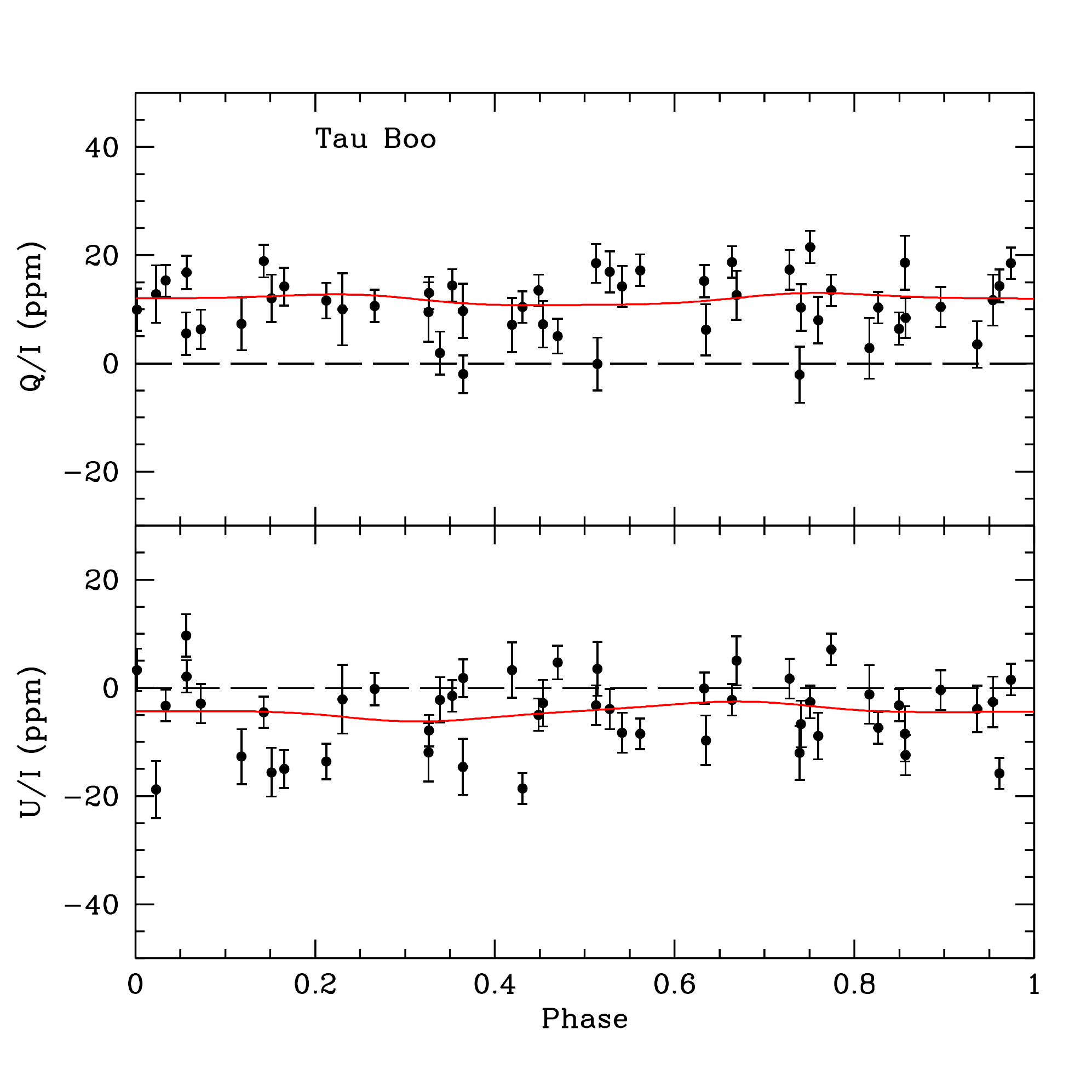}
    \includegraphics[width=8.8cm]{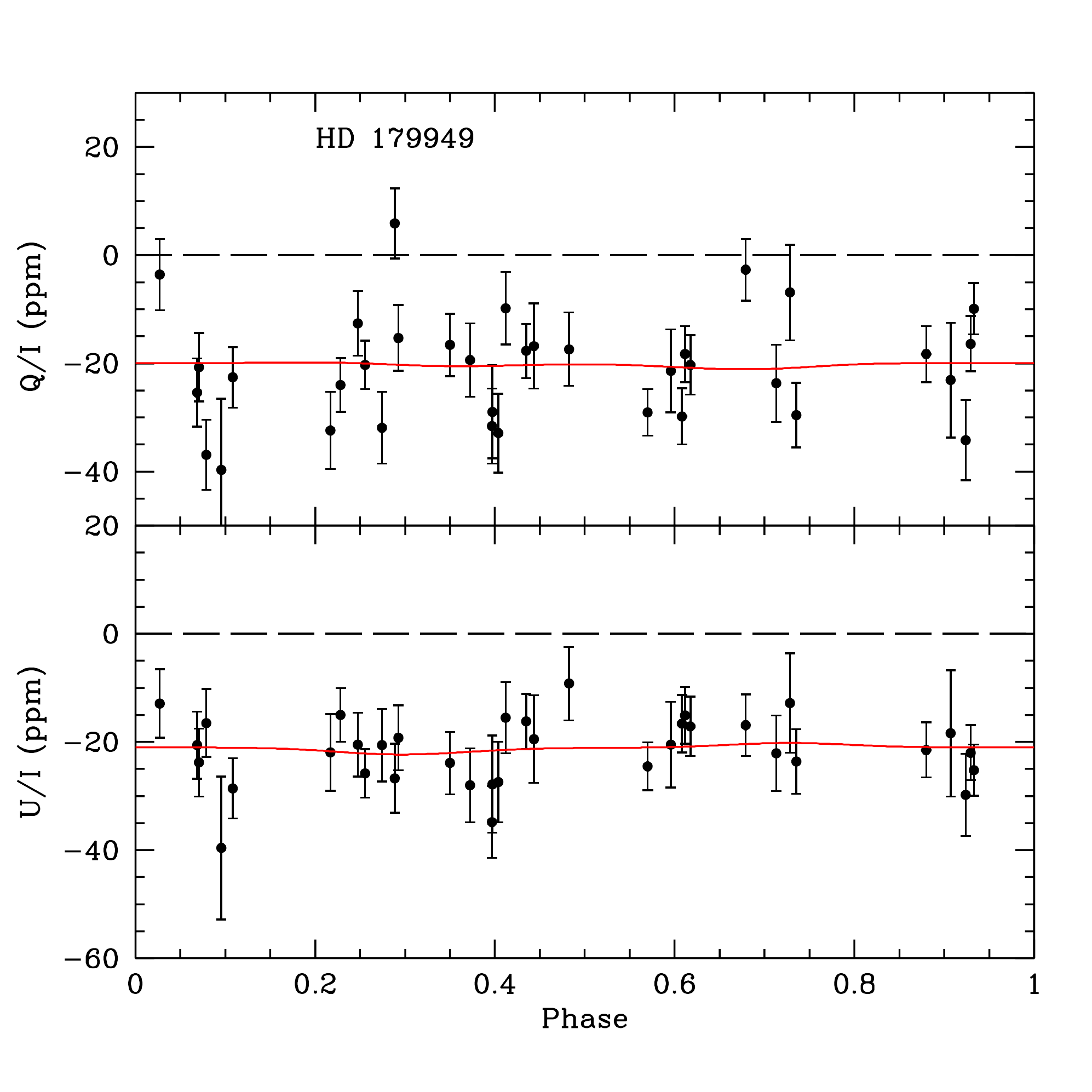}
    \includegraphics[width=8.8cm]{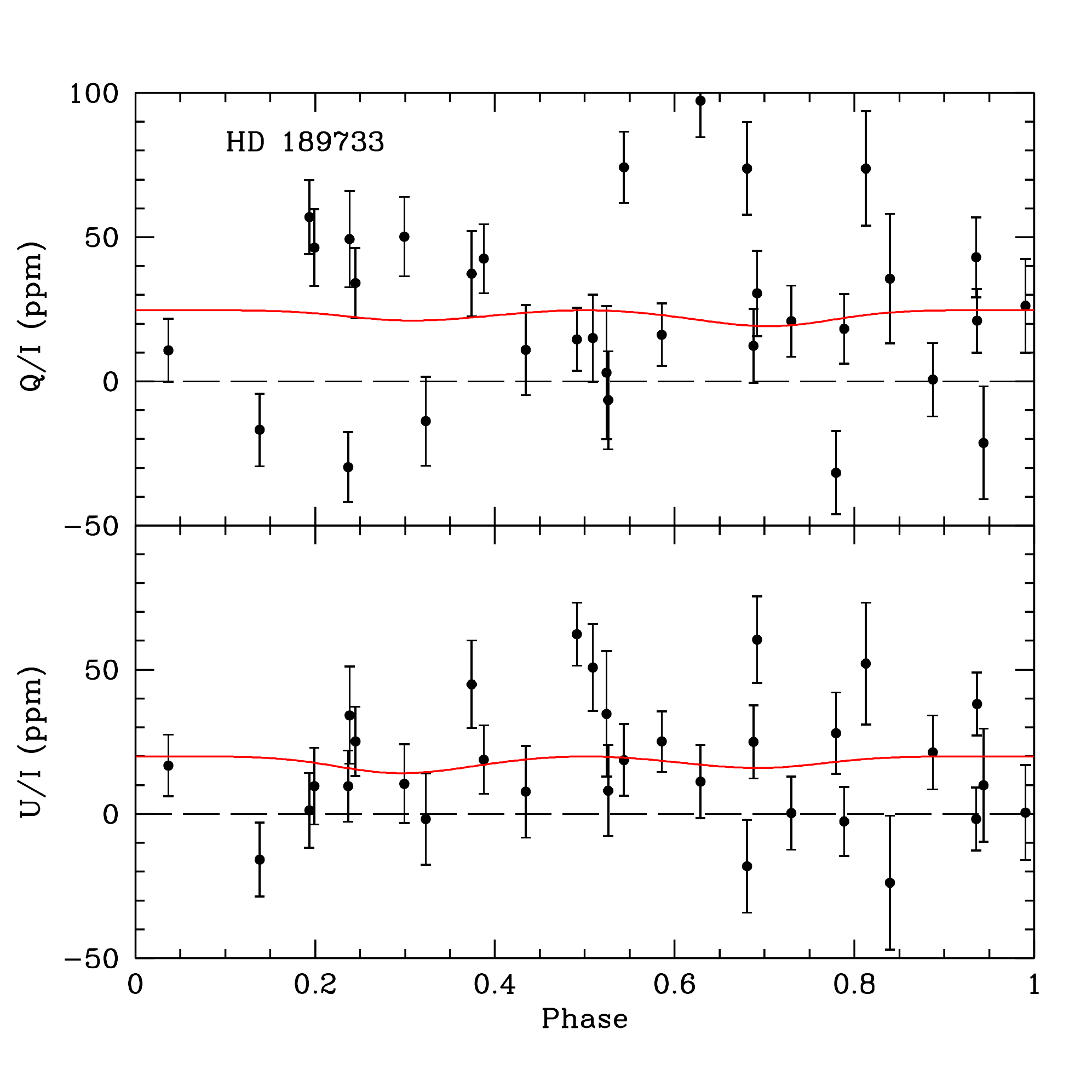}
    \includegraphics[width=8.8cm]{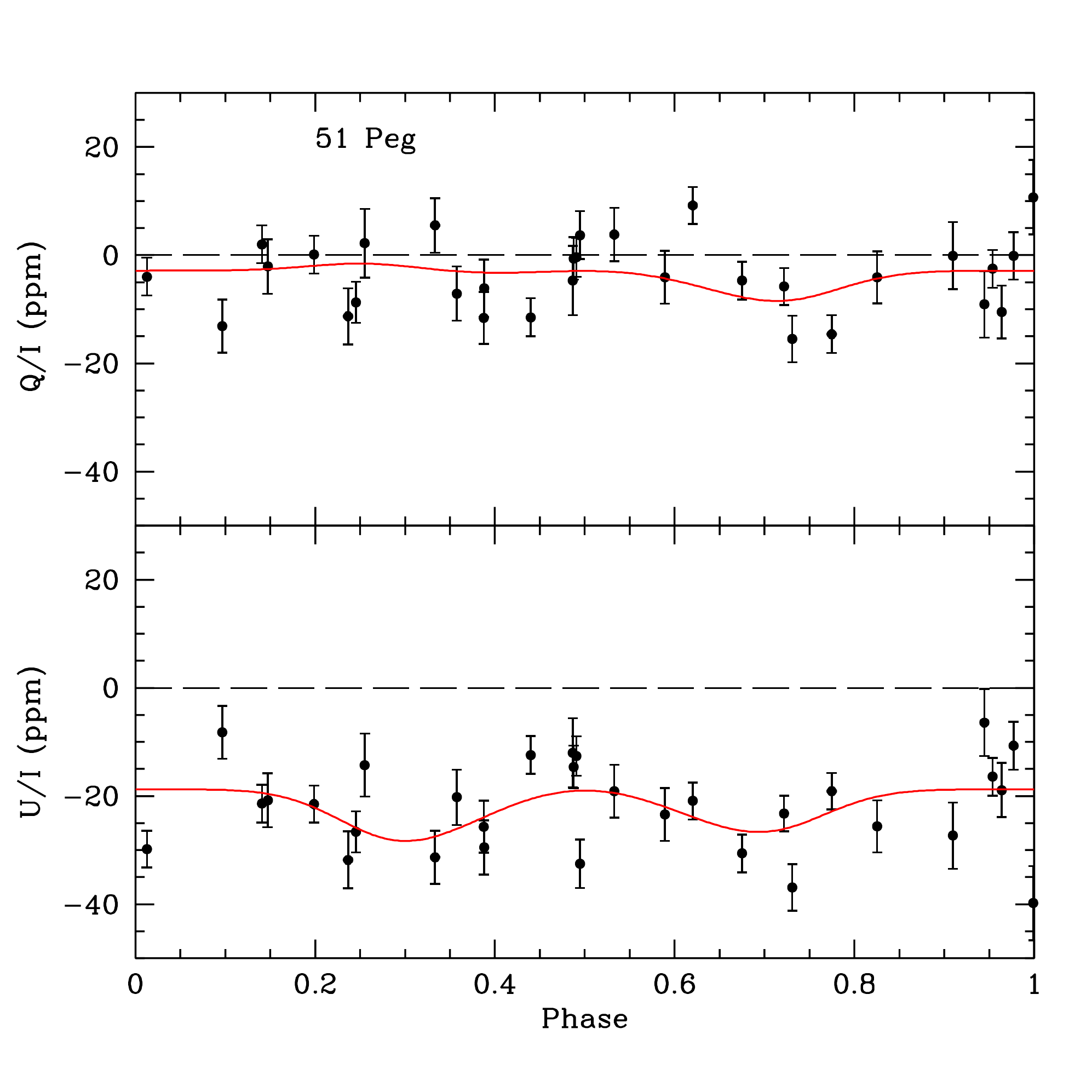}
    \caption{Polarization data plotted against orbital phase with fitted Rayleigh-Lambert models.}
    \label{fig:phase}
\end{figure*}

\subsection{Statistical Analysis}
\label{sec:stats}

\begin{table*}
\centering
\caption{Moment calculations.}
\tabcolsep 2 pt
\begin{tabular}{lc|cccccc|cccccc|c}
\hline
\hline
Object & n      & \multicolumn{6}{c|}{$Q/I$} & \multicolumn{6}{c|}{$U/I$} & $P$\\
    &           & Wt. Mean & Mean Err. & Std. Dev. & Err. Var. & Kurtosis & Skewness & Wt. Mean & Mean Err. & Std. Dev. & Err. Var. & Kurtosis & Skewness & Err. Var\\
\hline
$\tau$ Boo & 49 & \phantom{$-$}12.0 $\pm$ 0.5 & \phantom{0}3.9 & \phantom{0}5.6 & \phantom{0}4.0 & 0.155 & 2.678 & \phantom{0}$-$4.4 $\pm$ 0.5 & \phantom{0}3.9 & \phantom{0}6.8 & \phantom{0}5.6 & 0.051 & 2.364 & \phantom{0}6.9\\ 
HD~179949 & 36 & $-$20.3 $\pm$ 1.0 & \phantom{0}6.6 & 10.0 & \phantom{0}7.6 & 0.197 & 3.039 & $-$21.3 $\pm$ 1.0 & \phantom{0}6.6 & \phantom{0}6.2 & \phantom{0}0.0 & 0.344 & 3.571 & \phantom{0}7.6 \\
HD~189733 & 32 & \phantom{$-$}24.7 $\pm$ 2.4 & 14.4 & 31.0 & 27.4 & 0.039 & 2.637 & \phantom{$-$}18.1 $\pm$ 2.4 & 14.4 & 21.5 & 16.0 & 0.088 & 2.572 & 31.7\\
51 Peg & 31 & \phantom{0}$-$3.8 $\pm$ 0.8 & \phantom{0}4.6 & \phantom{0}6.6 & \phantom{0}4.8 & 0.031 & 2.423 & $-$21.9 $\pm$ 0.8 & \phantom{0}4.6 & \phantom{0}8.3 & \phantom{0}6.9 & 0.010 & 2.326 & \phantom{0}8.4 \\
\hline
\hline
\end{tabular}
\label{tab:stats}
\begin{flushleft}
Notes: Wt. Mean is the error weighted mean. Error variance (Err. Var.) is $\sqrt(x^2-e^2)$, where $x$ is the standard deviation (Std. Dev.) and $e$ the mean error (Mean Err.) of a set of measurements \mbox{($=0$ \textit{if} $x < e$)}; all of these quantities, along with the means are in ppm. None of the values for kurtosis or skewness are at all close to being significant at the 95~per~cent level, according to the tables of \citet{brooks94}. The final column in the table is the quadratic sum of the $Q/I$ and $U/I$ error variances, and represents the unaccounted for variability in the measurements, i.e. the potential signal level. \\
\end{flushleft}
\end{table*}

The moments of each data set, along with the `error variance' have been calculated and are reported in Table \ref{tab:stats}. The tables of \citet{brooks94} show that there is no significant skewness or kurtosis. This is not necessarily an indication that there is no intrinsic polarization, as such signals can have a Gaussian distribution. 

There are non-zero error variances for all of the systems. The quadratic sum of the error variances for Q/I and U/I give a value for P (final column in Table \ref{tab:stats}), that gives a potential signal level (assuming there are no other unaccounted for sources of variation). This gives 31~ppm for HD~189733, the next highest value is 8.4~ppm for 51~Peg which is a plausibly attributable to the planet. The other two systems have similar values for the error variance in P. The latter three signals are not strong. However, HD 179949 and 51 Peg are a little fainter than the stars used for instrumental precision determination \citep{bailey15, bailey20}, so non-intrinsic noise is also plausibly responsible. 

\subsection{Rayleigh-Lambert Model}

\label{sec:raylam}

We fit the corrected Q/I and U/I data points for each system with a Rayleigh-Lambert model \citep{seager00,wiktorowicz09} following the procedure described by \citet{bott18}. This is a simple analytic model for the expected polarization phase dependence of reflected light that assumes the planet reflects as a Lambert sphere and the polarization follows a Rayleigh scattering phase function. The resulting polarization variations closely resemble the result of more complete radiative transfer modelling of the planetary atmospheres for Rayleigh-like clouds \citep{bailey18,bott18}.

\begin{table}
    \centering
    \caption{Rayleigh-Lambert Model Fit Parameters}
    \begin{tabular}{lrrrr}
    Parameter & $\tau$ Boo & HD 179949 & HD 189733 & 51 Peg  \\ \hline
    $Z_q$ (ppm)  & 11.9 $\pm$ 0.8 & $-$20.0 $\pm$ 1.6 & 24.7 $\pm$ 5.3 & $-3.8 \pm 1.3$ \\
    $Z_u$ (ppm)  & $-4.4 \pm 1.1$ & $-21.0 \pm 1.3$   & 19.7 $\pm$ 6.0 & $-$18.4 $\pm$ 2.0 \\
    $P$ (ppm)    & 1.8 $\pm$ 1.7  & 1.4 $\pm$ 2.0     & 6.8 $\pm$ 12.0 & 11.2 $\pm$ 4.0 \\
    $PA$ (deg)   & 150 $\pm$ 45   & 134 $\pm$ 48      & 121 $\pm$ 51   & 157 $\pm$ 13.5   \\
    $i$ (deg)    & 45             &   67.7            &   85.5         &   76           \\
    \hline
    \end{tabular}
    \label{tab:rlambert}
\end{table}

\begin{figure}
    \centering
    \includegraphics[width=8.8cm]{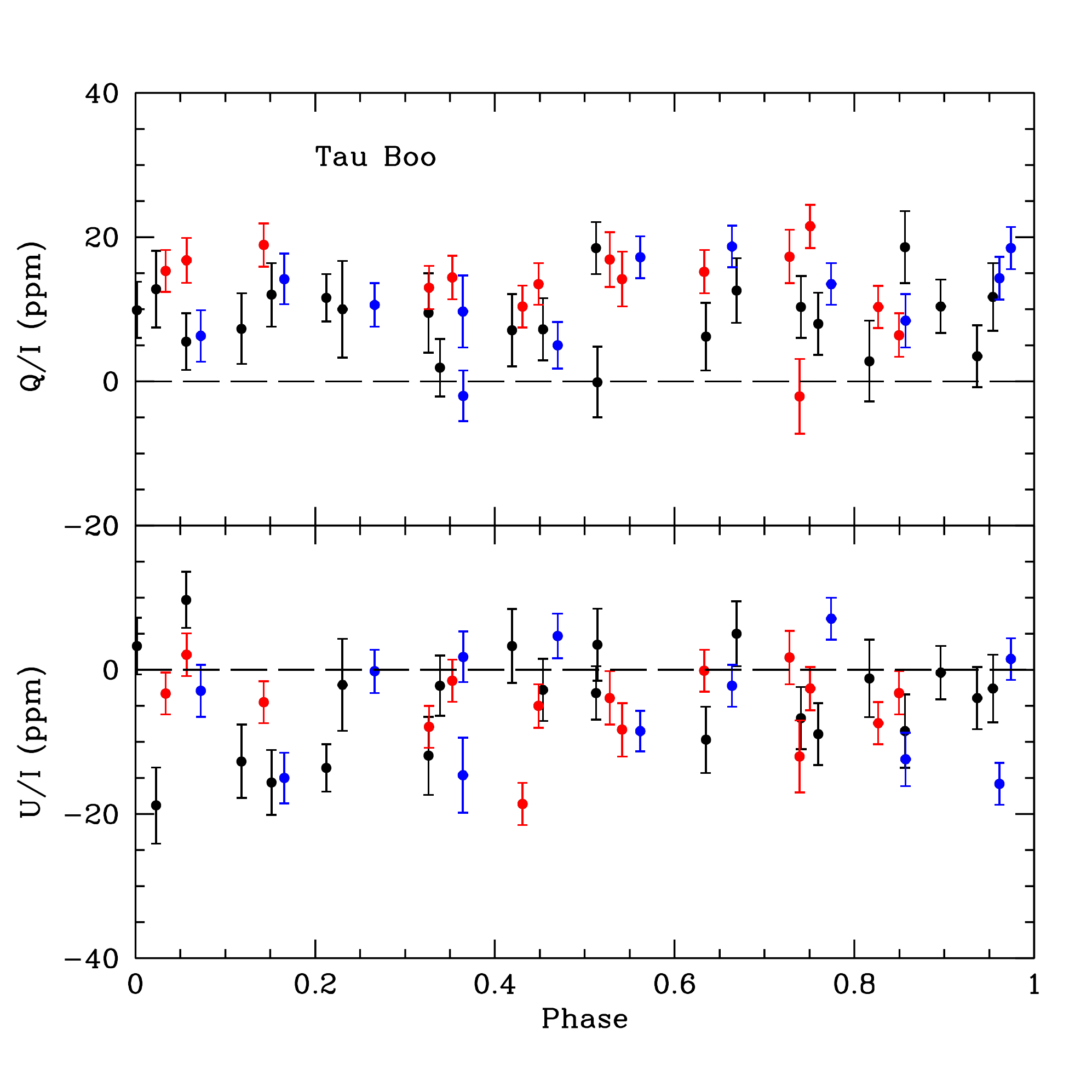}
    \caption{Corrected polarization data for $\tau$ Boo plotted against orbital phase. Red points are from the 2018MAR run, and blue points are from the 2018JUL run, with other runs plotted in black.}
    \label{fig:tauboo2}
\end{figure}

The results of the fits are shown in Table \ref{tab:rlambert}. There are four fitted parameters, the polarization offsets ($Z_q$ and $Z_u$, which allow for constant polarization from other sources, such as interstellar polarization), the polarization amplitude ($P$) and the position angle of the line of nodes of the orbit ($PA$). A fifth parameter, the inclination ($i$) is not fitted, but fixed at the values from Table \ref{tab:properties}. The fitting is performed with a Levenberg-Marquardt non-linear least-squares algorithm \citep{press92}. The errors quoted in table \ref{tab:rlambert} were obtained using the bootstrap method \citep{press92}. For each object we generated 10\,000 trial data sets by randomly selecting observations with replacement, so that each observation may be selected multiple times or not at all. We repeated the Rayleigh-Lambert fit to each of these trials and determined the errors from the standard deviation of the results. This gives a better result than the errors derived from the covariance matrix  of the fit, because it captures the full scatter in the data, including effects such as noise due to stellar activity, that may not be included in the formal errrors on the polarization measurements. In Fig. \ref{fig:phase} the corrected data points are plotted against orbital phase with the fitted model overlaid. 

It can be seen from the fitted polarization amplitudes ($P$) in Table \ref{tab:rlambert} that we do not detect significant reflected light polarization signals in $\tau$~Boo, HD~179949 and HD~189733. This is particularly clear when we note that there is a statistical bias for the fitted amplitude resulting from the fact that amplitudes are always positive \citep{bott18}. The bias is most noticeable when the fitted value is close to the error. 

In the case of $\tau$ Boo the fitted value of P = 1.8 $\pm$ 1.7 ppm is significantly lower than the value of 13 ppm predicted from Table \ref{tab:predict}. However, this is consistent with the limit on the geometric albedo for this planet ($<$0.12) reported by \citet{hoeijmakers18}. Using the model results in Figure 14 of \citep{bailey18} scaled for $\tau$ Boo as described in section \ref{sec:objects} we find that a limit on geometric albedo of 0.12 corresponds to a polarization amplitude of less than $\sim$ 4 ppm, consistent with what we observe. We note that as explained by \citet{bailey18} it is not possible to use an upper limit on observed polarization to give an upper limit on the geometric albedo of a planet without assumptions about the nature of the scattering particles. For example, Venus is a planet with a high geometric albedo but a relatively low polarization. However, a low geometric albedo will certainly result in a low polarization.

$\tau$ Boo differs from the other three systems in that the rotation of the star is believed to be synchronized with the orbital period, so that polarization due to stellar activity could also vary periodically with orbital phase. However since the magnetic field and spot patterns change over long periods we would not expect such variations to be coherent over the 3 years covered by our observations. In Fig. \ref{fig:tauboo2} we show the phase variability highlighting the data from individual observing runs. If stellar activity was contributing significantly to the broad-band polarization we might see phase dependent signals in individual short runs. There is at best only weak evidence for this shown in Fig. \ref{fig:tauboo2}. 

For HD 179949 the fitted amplitude of $P$ = 1.4 $\pm$ 2.0 ppm is not significant. If there is a reflected light signal in this object it will require substantially improved data to detect it.

For HD 189733 the fitted value of $P$ = 6.8 $\pm$ 12.0 ppm has a large error, only slightly lower than that determined by \citet{bott16} from a much smaller data set. It is apparent from the scatter seen in figures \ref{fig:runs} and \ref{fig:phase} that there is substantial polarization variability in this object that does not follow the pattern expected for reflected light from the planet. As discussed below we attribute this polarization to magnetic activity of the host star. The resulting large error in the fitted polarization amplitude means we cannot reach any useful conclusion on the presence or absence of reflected light polarization. A signal level of $\sim$20 ppm that is consistent with predictions \citep{bailey18} and with the \citet{evans13} reflected light observation cannot be ruled out.

\begin{figure}
    \centering
    \includegraphics[width=8.8cm]{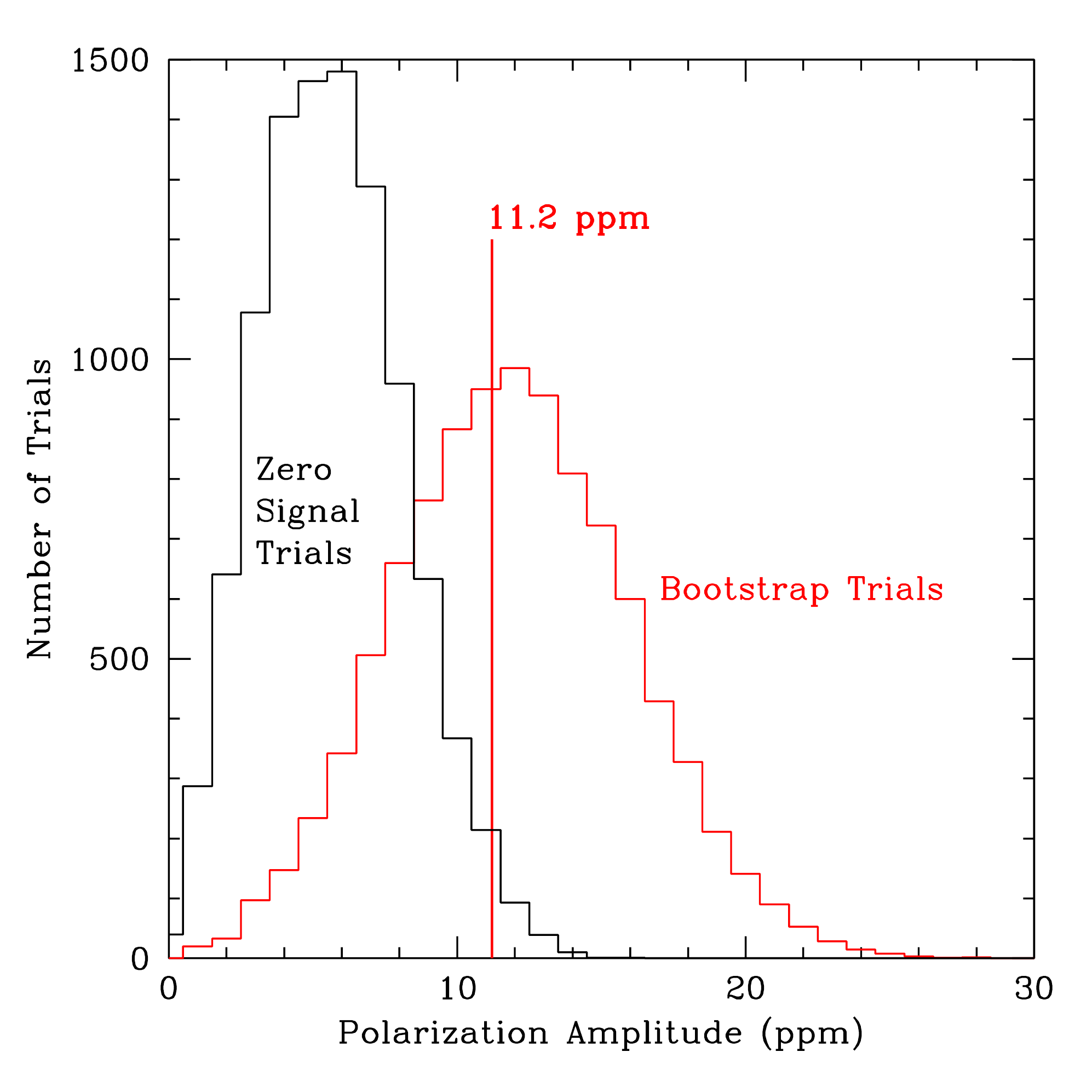}
    \caption{Statistical analysis of the possible reflected light signal in the Rayleigh-Lambert fit to the 51 Peg polarization data. The value directly fitted to the observations is at an amplitude of 11.2 ppm. The red histogram shows the result of 10\,000 bootstrap trials generated from the original data as explained in the text. The black histogram shows the amplitudes fitted to 10\,000 simulated data sets with zero reflected light signal and the same phasing and error properties as the observations. The amplitude of 11.2 ppm is exceeded by 1.9 per cent of these zero signal trials.}
    \label{fig:boothist}
\end{figure}

\subsection{Possible Reflected Light from 51 Peg b}

The Rayleigh-Lambert fit to the 51 Peg polarization data shows a signal of the form expected for reflected light with an amplitude of 11.2 ppm. This is comparable with the expected amplitude given in Table \ref{tab:predict} although it does not require the more extreme planetary radius of 1.9 $R_{\rm Jup}$. The fit was done for an inclination of 76$\degr$ but a similar fit is also obtained at the extremes of the range given in Table \ref{tab:properties} (70$\degr$ and 82$\degr$). 51~Peg is the least active of the four stars observed, so we are unlikely to be seeing polarization due to stellar activity. It is an object that also has a reported reflected light detection from spectroscopy \citep{martins15}.

 The histogram of amplitude values resulting from the bootstrap analysis (see Section \ref{sec:raylam}) of the data is shown in Fig. \ref{fig:boothist} and  provides a measure of the statistical uncertaintly on the measured value of 11.2 ppm.  As a further test we generated 10\,000 simulated data sets with zero reflected light signal, and the same phasing and error properties as the actual observations. We fitted these data sets with Rayleigh-Lambert models in the same way as the actual observations. The histogram of amplitudes is also shown in Fig. \ref{fig:boothist}. It can be seen that, although there is zero signal in these data sets, the histogram is centered on about 5.5 ppm, an indication of the bias in amplitude estimates due to the fact that amplitude can only be positive. From this histogram we find that the amplitude of 11.2 ppm is exceeded by 1.9\% of the zero signal random trials, providing an estimate of the false alarm probability for this reflected light signal. This figure of 1.9\% may be slightly underestimated if there are additional sources of noise not accounted for in the formal errors of the data points, such as noise due to stellar activity.

Based on the bootstrap analysis as given in Table \ref{tab:rlambert} the polarization amplitude is determined to 2.8$\sigma$ and the reflected light signal has a false alarm probability of 1.9 per cent as described above. These results are promising but are not sufficient to claim a conclusive detection of a reflected light polarization signal. More extensive observations or improved precision will be needed to confirm the presence of polarized reflected light.

\subsection{Polarization due to Host Star Magnetic Activity}

Most studies of polarization in hot Jupiter systems have assumed that the host star is unpolarized and thus any polarization variability will be due to light reflected from the planet. However, many exoplanet host stars are magnetically active and this may contribute to the polarization. \citet{cotton17b} investigated the broad band linear polarization of dwarfs of spectral types F, G and K, and found that active dwarfs showed polarization at levels up to $\sim$45 ppm.  \citet{cotton19a} monitored one of these active dwarfs ($\xi$ Boo A) and found polarization varying over the stellar rotation period with a fixed phase offset from the magnetic field as determined from circular spectropolarimetry. The polarization was larger for stars with larger magnetic fields ($\sim$ 4 ppm G$^{-1}$ according to \citealt{cotton19a}).

The likely cause of broad band linear polarization is thought to be differential saturation\footnote{Sometimes referred to in the literature as ``magnetic intensification''.} \citep{leroy89,leroy90} of Zeeman-split spectral lines in the stars' global magnetic field \citep{cotton19a}. In a transverse magnetic field spectral lines are split into three components, a central $\pi$ component polarized parallel to the field, and red and blue shifted $\sigma$ components polarized perpendicular to the field. The $\pi$ component has twice the strength of the $\sigma$ components, and thus in the weak-line case there is no net polarization averaged over all the components. However, since the $\pi$ components are stronger than the $\sigma$ components they are more affected by saturation, and in a stellar atmosphere with many overlapping spectral lines this leads to a small net broad band polarization.

Another possible polarization mechanism is Rayleigh scattering in the stellar atmosphere with the symmetry of a spherical star being broken by the presence of starspots. However this effect is found to be smaller than the differential saturation effect \citep{saar93,kostogryz15}.

The variable polarization we see in HD 189733 (Table \ref{tab:stats}, Figures \ref{fig:runs} and \ref{fig:phase}) is likely to be the result of host star activity. The variations do not correlate with orbital phase. HD~189733 is the most active host star in our sample and has the strongest magnetic field. With a mean field of 32--42 G \citep{fares17} the apparent variability of $\sim\pm50$ ppm seen in Fig. \ref{fig:phase} is reasonably consistent with results for other active dwarfs \citep{cotton17b,cotton19a}. Polarization due to stellar activity may, in part, explain the discrepancies in polarization results reported in past studies of HD~189733 \citep{berdyugina08,wiktorowicz09,berdyugina11,wiktorowicz15,bott16}. More specifically, the host star activity, as well as the effects of a Saharan dust event on some of the observations \citep{bott16}, may explain the $\sim$100 ppm polarization amplitude reported by \citet{berdyugina11} for HD 189733. This amplitude is too large to be reflected light from the planet \citep{lucas09,bailey18} and has not been seen in subsequent studies \citep{wiktorowicz09,wiktorowicz15,bott16} or in the results presented here.

We have looked for periodic polarization variations of HD~189733 over the stellar rotation period without success. This is not surprising in view of the differential rotation and changing magnetic field pattern (Section \ref{sec:hd189733}) and the length of our observing runs which is typically less than the $\sim$12 day rotation period. We cannot expect polarization variations to remain coherent over the $\sim$4 years covered by the observations. 

HD~179949 and $\tau$~Boo are also known to be magnetically active stars as discussed in Section \ref{sec:tauboo} and \ref{sec:hd179949}. Photometry of $\tau$ Boo has shown variations attributed to an active spot \citep{walker08}. While it is possible that activity contributes to the observations of polarization, the amplitude of any changes in both of these systems are much less than seen in HD 189733. The main reason for this is likely to be the much lower magnetic field (see Table \ref{tab:properties}).

\subsection{Interstellar Polarization}

The polarization of the exoplanet systems will be a combination of any intrinsic polarization and interstellar polarization. If the interstellar polarization can be determined independently from observations of the system, this can provide useful information on the likely (constant/mean) intrinsic polarization level. Observing nearby intrinsically unpolarized stars is a common way of gauging the magnitude and orientation of interstellar polarization for a target \citep{clarke10}. We have previously found stars with spectral types ranging from A to early K to be the least intrinsically polarized \citep{bailey10, cotton16a, cotton16b}. Such stars are a fair probe of the nearby interstellar medium so long as debris-disk hosts \citep{cotton17b, vandeportal19}, particularly active stars \citep{cotton19a} or early A-type rapid rotators \citep{cotton17a, bailey20b} are avoided. 

\begin{figure*}
\includegraphics[clip, trim={2.7cm 0cm 2.5cm 1cm}, width=\textwidth]{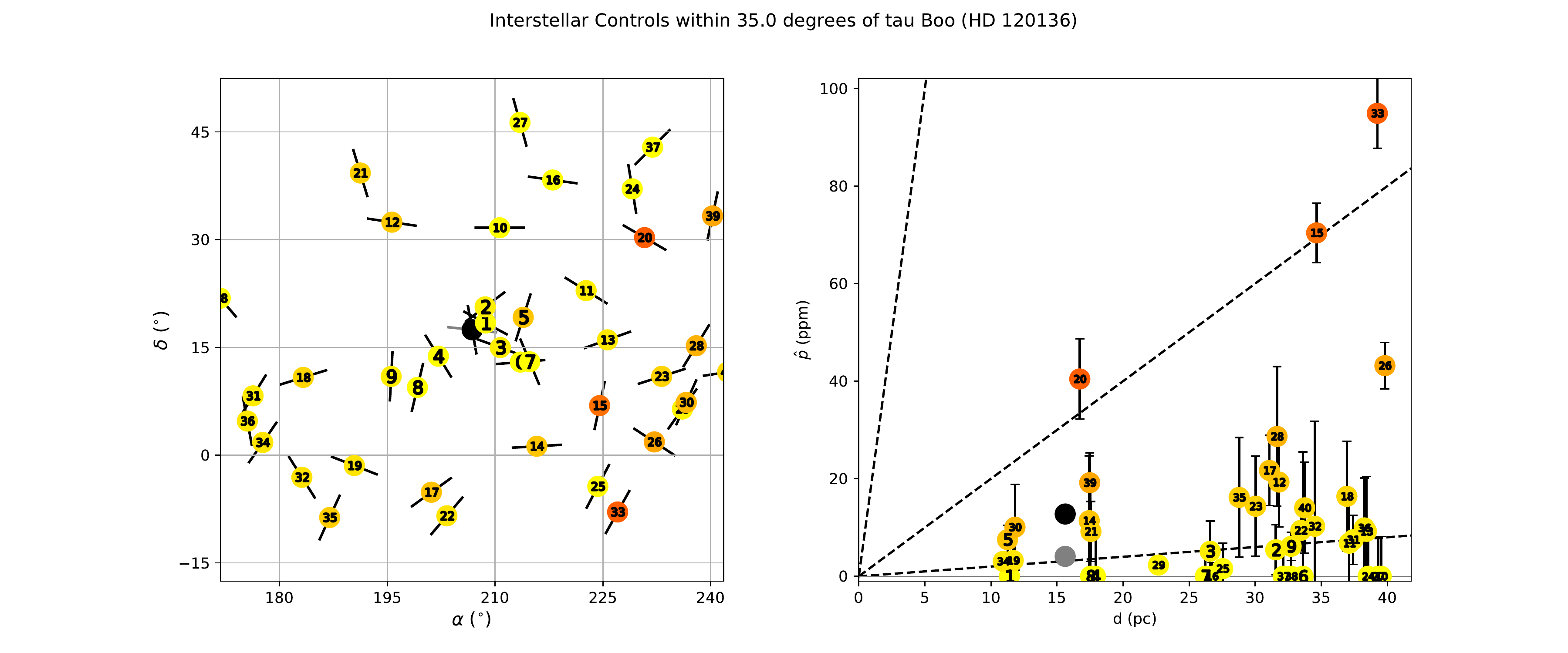}
\caption{A map (left) and p vs d plot (right) of interstellar control stars within 35$^\circ$ of \textbf{$\tau$~Boo}. Interstellar PA ($\theta$) is indicated on the map by the black pseudo-vectors; and defined as the angle North through East, i.e. increasing in a clockwise direction with vertical being 0$^\circ$. The controls are colour coded in terms of $\hat{p}/d$ and numbered in order of their angular separation from $\tau$~Boo; they are: 1:~HD~121370, 2:~HD~121320, 3:~HD~122320, 4:~HD~117176, 5:~HD~124897, 6:~HD~124570, 7:~HD~125451, 8:~HD~115383, 9:~HD~113226, 10:~HD~122652, 11:~HD~131042, 12:~HD~113319, 13:~HD~133161, 14:~HD~126053, 15:~HD~132307, 16:~HD~127762, 17:~HD~116568, 18:~HD~106210, 19:~HD~110379J, 20:~HD~137107A, 21:~HD~110897, 22:~HD~117860, 23:~HD~138573, 24:~HD~135891, 25:~HD~132052, 26:~HD~137898, 27:~HD~124694, 28:~HD~142093, 29:~HD~140573, 30:~HD~141004, 31:~HD~102124, 32:~HD~106116, 33:~HD~134088, 34:~HD~102870, 35:~HD~108510, 36:~HD~101690, 37:~HD~138004, 38:~HD~99505, 39:~HD~143761, 40:~HD~145229. In the p vs d plot dashed lines corresponding to $\hat{p}/d$ values of 0.2, 2.0 and 20.0~ppm/pc are given as guides. The grey data-point is derived from the interstellar model in \citep{cotton17b} and the black data-point represents our best-fit interstellar values for $\tau$~Boo (converted to 450~nm).}
 \label{fig:is_tau_Boo}
\end{figure*}

\begin{figure*}
\includegraphics[clip, trim={2.7cm 0cm 2.5cm 1cm}, width=\textwidth]{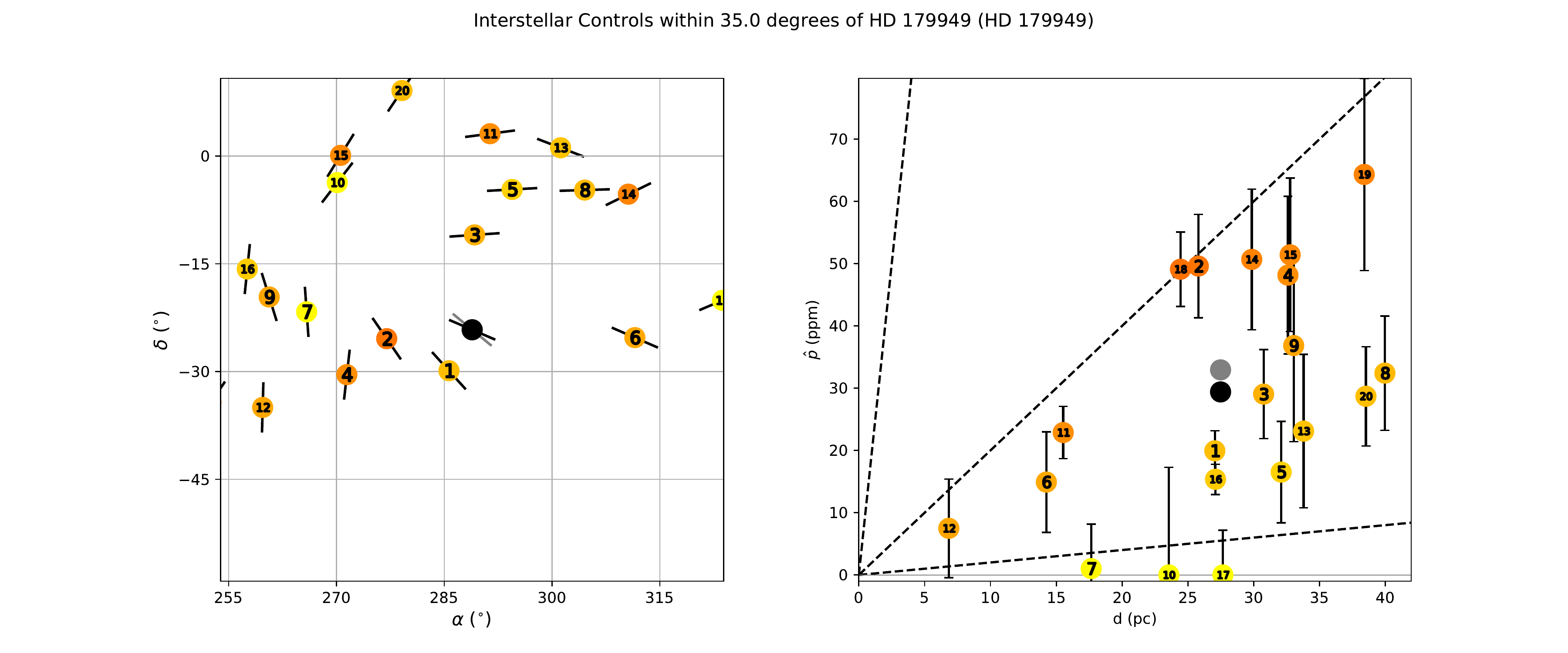}
\caption{A map (left) and p vs d plot (right) of interstellar control stars within 35$^\circ$ of \textbf{HD~179949}. Interstellar PA ($\theta$) is indicated on the map by the black pseudo-vectors; and defined as the angle North through East, i.e. increasing in a clockwise direction with vertical being 0$^\circ$. The controls are colour coded in terms of $\hat{p}/d$ and numbered in order of their angular separation from HD~179949; they are: 1:~HD~176687, 2:~HD~169916, 3:~HD~180409, 4:~HD~165135, 5:~HD~185124, 6:~HD~197692, 7:~HD~160915, 8:~HD~193017, 9:~HD~157172, 10:~HD~164259, 11:~HD~182640, 12:~HD~156384, 13:~HD~190412, 14:~HD~197210, 15:~HD~164651, 16:~HD~155125, 17:~HD~205289, 18:~HD~151680, 19:~HD~151504, 20:~HD~171802. In the p vs d plot dashed lines corresponding to $\hat{p}/d$ values of 0.2, 2.0 and 20.0~ppm/pc are given as guides. The grey data-point is derived from the interstellar model in \citep{cotton17b} and the black data-point represents our best-fit interstellar values for HD~179949 (converted to 450~nm).}
 \label{fig:is_HD179949}
\end{figure*}

\begin{figure*}
\includegraphics[clip, trim={2.7cm 0cm 2.5cm 1cm}, width=\textwidth]{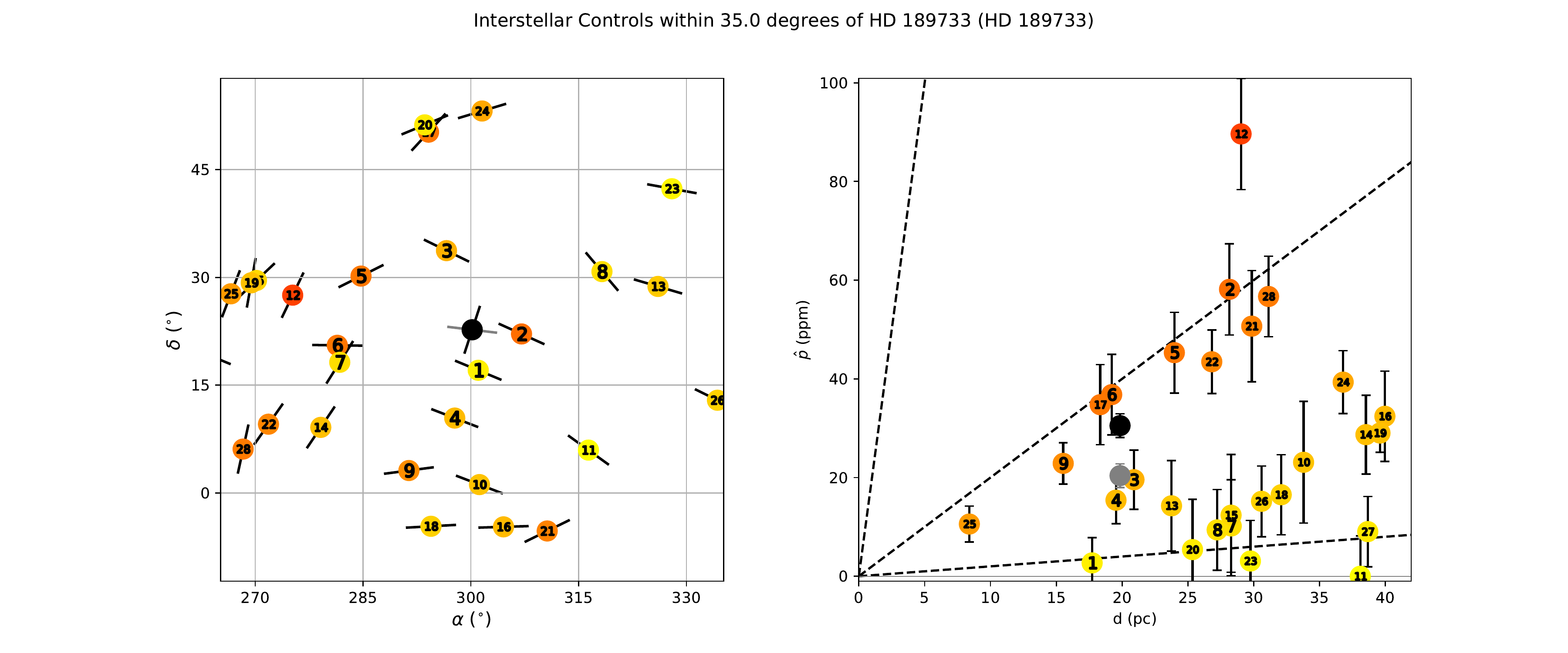}
\caption{A map (left) and p vs d plot (right) of interstellar control stars within 35$^\circ$ of \textbf{HD~189733}. Interstellar PA ($\theta$) is indicated on the map by the black pseudo-vectors; and defined as the angle North through East, i.e. increasing in a clockwise direction with vertical being 0$^\circ$. The controls are colour coded in terms of $\hat{p}/d$ and numbered in order of their angular separation from HD~189733; they are: 1:~HD~190406, 2:~HD~195034, 3:~HD~187013, 4:~HD~187691, 5:~HD~176377, 6:~HD~173667, 7:~HD~173880, 8:~HD~202108, 9:~HD~182640, 10:~HD~190412, 11:~HD~200790, 12:~HD~168874, 13:~HD~206826, 14:~HD~171802, 15:~HD~164595, 16:~HD~193017, 17:~HD~185395, 18:~HD~185124, 19:~HD~163993, 20:~HD~184960, 21:~HD~197210, 22:~HD~165777, 23:~HD~207966A, 24:~HD~191195, 25:~HD~161797, 26:~HD~211476, 27:~HD~159332, 28:~HD~162917. In the p vs d plot dashed lines corresponding to $\hat{p}/d$ values of 0.2, 2.0 and 20.0~ppm/pc are given as guides. The grey data-point is derived from the interstellar model in \citep{cotton17b} and the black data-point represents our best-fit interstellar values for HD~189733 (converted to 450~nm).}
 \label{fig:is_HD189733}
\end{figure*}

A number of control stars near our target systems are to be found in the recently released catalogue of \citet{piirola20} which collates observations made with the DIPOL-2 polarimeter \citep{piirola14} -- from which we have removed a few known debris disk hosts. More controls are to be found amongst the `ordinary FGK dwarfs' and the \textit{Interstellar List} in \citet{cotton17b}, as well as amongst the controls in \citet{bailey20b} and \citet{cotton20a}. Except for a few of the Interstellar List originally observed with PlanetPol \citep{bailey10}, all the controls have been observed using combined Johnson B, V and R filters (the data from \citealp{piirola20}) or the SDSS g$^\prime$ filter (everything else). The different filters and spectral types results in a range of $\lambda_{\rm eff}$ values. For the purpose of mapping the interstellar polarization in figures \ref{fig:is_tau_Boo}, \ref{fig:is_HD179949}, \ref{fig:is_HD189733} and \ref{fig:is_51_Peg}, we use the Serkowski law \citep{serkowski75} with $\lambda_{max}$ equal to 470~nm (the value found by \citealp{marshall16}) to convert the control data to the value expected for 450~nm.  The wavelength of maximum interstellar polarization as near to the Sun as our targets is not well defined. However, a number of studies \citep{marshall16, cotton19b, bailey20b} indicate it is near this figure. 

\subsubsection{\texorpdfstring{$\tau$ Bo\"otis}{tau Bootis}}

There is no pattern in $\theta$ of the control stars near $\tau$~Boo. However, the polarization of $\tau$~Boo is a little higher than the nearest control stars -- those within 15$^{\circ}$ -- the polarization of which is very low\footnote{As an aside we note that the control which is the greatest outlier, HD~137107A a G2~V star, is separated from its G2~V companion by only 1$\arcsec$, and on this basis warrants further investigation}. This may indicate a small intrinsic polarization of $\sim$10~ppm. This is consistent with the level of stellar activity displayed by $\tau$~Boo \citep{cotton19a}, which may be expected to produce polarization that will not completely average out over a rotation phase cycle.

\subsubsection{HD~179949}

The polarization of HD~179949 is in line with those of the nearest controls -- which have a wider range of values than in the case of $\tau$~Boo -- and is very close to that predicted by the model in \citet{cotton17b}; $\theta$ also aligns very well with the prediction (which is based on the values of the control stars). 

\subsubsection{HD~189733}

Although the magnitude of polarization of HD~189733 is reasonably consistent with that of nearby stars, $\theta$ is almost perpendicular to those of nearest stars. It is worth noting that \citet{bott16}, working with just the data from 2014 and 2015, found a very different baseline polarization level which they pointed out was much higher than expected. Clearly HD~189733's variable activity is significantly affecting these determinations. 

\subsubsection{51~Pegasi}

\begin{figure*}
\includegraphics[clip, trim={2.7cm 0cm 2.5cm 1cm}, width=\textwidth]{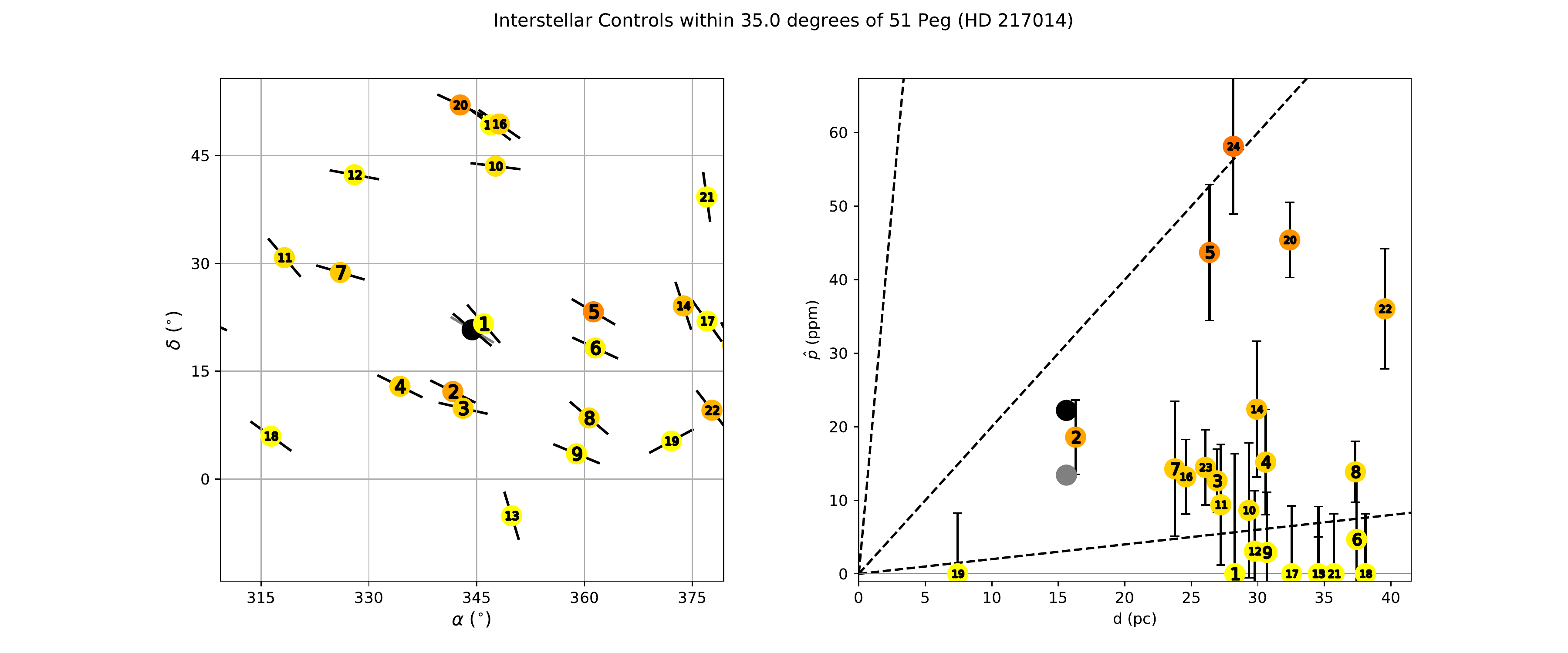}
\caption{A map (left) and p vs d plot (right) of interstellar control stars within 35$^\circ$ of \textbf{51~Peg}. Interstellar PA ($\theta$) is indicated on the map by the black pseudo-vectors; and defined as the angle North through East, i.e. increasing in a clockwise direction with vertical being 0$^\circ$. The controls are colour coded in terms of $\hat{p}/d$ and numbered in order of their angular separation from 51~Peg; they are: 1:~HD~217924, 2:~HD~215648, 3:~HD~216385, 4:~HD~211476, 5:~HD~225261, 6:~HD~101, 7:~HD~206826, 8:~HD~225003, 9:~HD~224156, 10:~HD~218804, 11:~HD~202108, 12:~HD~207966A, 13:~HD~219877, 14:~HD~5294, 15:~HD~218470, 16:~HD~219080, 17:~HD~6715, 18:~HD~200790, 19:~HD~4628, 20:~HD~216275, 21:~HD~6664, 22:~HD~7047, 23:~HD~8262, 24:~HD~195034. In the p vs d plot dashed lines corresponding to $\hat{p}/d$ values of 0.2, 2.0 and 20.0~ppm/pc are given as guides. The grey data-point is derived from the interstellar model in \citep{cotton17b} and the black data-point represents our best-fit interstellar values for 51~Peg (converted to 450~nm).}
 \label{fig:is_51_Peg}
\end{figure*}

In Fig. \ref{fig:is_51_Peg} it is the mean polarization for 51~Peg as per Table \ref{tab:stats} that is plotted as the black circle -- as it is in corresponding figures for the other targets. The magnitude of this polarization is similar to that expected based on the nearest stars and the model represented by the grey circle; the value of $\theta$ also matches very closely that expected from interstellar polarization. The values from Table \ref{tab:rlambert} are very similar to those from Table \ref{tab:stats}, and if substituted would make no appreciable difference. 

\section{Conclusions}
\label{sec:conclude}

We have presented high-precision linear polarization observations of four bright hot Jupiter systems. We analyse the data to search for the polarization signal expected for reflected light from the planet. Only one of the four systems, 51~Peg, shows evidence for a reflected light signal of the form expected. The result has a 2.8$\sigma$ significance, and a false alarm probability of about 1.9 per cent. Further observations will be required to reach a definitive conclusion on the presence of reflected light polarization. The observed polarization signal is consistent with the reflected light detection from spectroscopy \citep{martins15}, although does not require the extremely large planetary radius that was suggested to explain that result. 51 Peg is the least active of the four stars observed.

HD~189733 shows substantial evidence for variable polarization, but there is no evidence for any signal repeating over the orbital period. The polarization is interpreted as being due to activity of the host star and is consistent with the broad-band polarization levels reported for other active dwarfs \citep{cotton17b,cotton19a}. HD~189733 has a much larger magnetic field than the three other host stars, and the polarization can be attributed to differential saturation \citep{leroy89,leroy90} in the global magnetic field. The variable host star polarization masks any signal due to reflected light from the planet. A reflected light polarization signal at a level $\sim$20 ppm that would be consistent with the reflected light reported by \citet{evans13} cannot be excluded.

$\tau$~Boo shows no evidence for a significant polarized light signal. This is consistent with the result from \citet{hoeijmakers18} that the geometric albedo is very low. We also see no significant polarization signal in HD~179949.

Comparing the mean polarization of each system with nearby unpolarized stars, we find evidence for a constant/mean intrinsic component of polarization, probably as a result of stellar activity in the $\tau$~Boo and HD~189733 systems. For the 51~Peg and HD~179949 systems the mean polarization is consistent with being interstellar polarization.

The results show that detection of polarized reflected light from hot Jupiter systems is difficult with current techniques. One significant issue is polarization due to host star activity. Active stars, and particularly those with strong magnetic fields, need to be avoided. In this respect, HD 189733, which has been one of the most studied targets for polarimetry \citep{berdyugina08,wiktorowicz09,berdyugina11,wiktorowicz15,bott16}, turns out to be a poor choice.

Even when activity is not such a serious issue, some hot Jupiters are known to have very low geometric albedos \citep[e.g.][]{kipping11,mocnik18} and this makes detection of polarization unlikely. One of our systems, $\tau$~Boo has now been shown to have a low albedo \citep{hoeijmakers18}.

Even in the best cases, the sensitivity of current polarimeters is marginal for this purpose and improvements in instrumental performance are likely to be needed to make further progress. We need to improve the precision of individual polarization measurements and extend the sensitivity to fainter objects to increase the range of targets available for study. There are systematic effects that can limit the accuracy of observations with current instrument designs \citep{berdyugin18}. Calibration of the data for instrumental and telescope polarization contributions needs to be improved, and this requires a better understanding of the polarization properties of nearby stars.

\section*{Acknowledgements}

This work was supported by the Australian Research Council through Discovery Projects grant DP160103231. The research has been supported by the Ministry of Science and Technology of Taiwan under grants MOST107-2119-M-001-031-MY3, MOST107-2119-M-001-031-MY3, and MOST109-2112-M-001-036-MY3, and Academia Sinica under grant AS-IA-106-M03. We thank the staff of the Anglo-Australian Telescope for their support of our observing. Nicholas Borsato, Behrooz Karamiqucham, Fiona Lewis, Shannon Melrose and Daniela Opitz assisted with some of the AAT observations.


\section*{Data Availability}

The original data used in this study is provided in reduced form in Tables \ref{tab:tauboo} to \ref{tab:51peg}. Raw data files can be provided on request by the authors.



\bibliographystyle{mnras}
\bibliography{exoplanet_pol} 







\bsp	
\label{lastpage}
\end{document}